\documentclass[twocolumn,3p,times,procedia]{elsarticle}

\usepackage{ecrc}
\usepackage{graphicx}
\usepackage{balance}
\usepackage{array,threeparttable}
\usepackage{url}

\usepackage{soul}

\usepackage{xcolor}

\usepackage{caption}
\usepackage{subcaption}

\volume{00}

\firstpage{1}

\runauth{}

\jid{procs}

\CopyrightLine{2022}{Published by Elsevier Ltd.}

\usepackage{amssymb}
\usepackage{amsmath}
\usepackage{multicol}
\usepackage{algorithm}
\usepackage{algpseudocode}

\usepackage[figuresright]{rotating}
\usepackage{bm}
\usepackage{caption}
\usepackage{titlesec}
\titleformat*{\section}{\small \bf}
\titleformat*{\subsection}{\small \em}
\titleformat*{\subsubsection}{\small \em}

\usepackage{fancyhdr}
\fancyheadoffset[RO,EL]{0pt}
\usepackage{geometry}
\begin{document}\small
\begin{frontmatter}




\dochead{}
\title{
\begin{flushleft}
{\LARGE A deep-learning-based MAC for integrating channel access, rate adaptation and channel switch} 
\end{flushleft}
}
 %

\author[]{ \leftline {Jiantao Xin $^a$, Wei Xu $^b$, Bin Cao $^c$, Taotao Wang $^{a,}$$^*$, Shengli Zhang $^a$}}

\address{ \leftline {$^a$ College of Electronic and Information Engineering, Shenzhen University, Shenzhen 518060, China}
  \leftline {$^b$ Computer Science Department, New York University 251 Mercer Street, New York, NY 10012, USA}
  \leftline {$^c$ The State Key Laboratory of Networking and Switching Technology, Beijing University of Posts and Telecommunications, Beijing 100876, China}
}

\cortext[]{Corresponding author}
\ead{xinjiantao2017@email.szu.edu.cn (Jiantao Xin)}
\ead{wx317@nyu.edu (Wei Xu)}
\ead{caobin@bupt.edu.cn (Bin Cao)}
\ead{ttwang@szu.edu.cn (Taotao Wang)}
\ead{zsl@szu.edu.cn (Shengli Zhang)}

\begin{abstract}

With increasing density and heterogeneity in unlicensed wireless networks, traditional MAC protocols, such as carrier-sense multiple access with collision avoidance (CSMA/CA) in Wi-Fi networks, are experiencing performance degradation. This is manifested in increased collisions and extended backoff times, leading to diminished spectrum efficiency and protocol coordination. Addressing these issues, this paper proposes a deep-learning-based MAC paradigm, dubbed DL-MAC, which leverages spectrum sensing data readily available from energy detection modules in wireless devices to achieve the MAC functionalities of channel access, rate adaptation and channel switch. First, we utilize DL-MAC to realize a joint design of channel access and rate adaptation. Subsequently, we integrate the capability of channel switch into DL-MAC, enhancing its functionality from single-channel to multi-channel operation. Specifically, the DL-MAC protocol incorporates a deep neural network (DNN) for channel selection and a recurrent neural network (RNN) for the joint design of channel access and rate adaptation. We conducted real-world data collection within the 2.4 GHz frequency band to validate the effectiveness of DL-MAC, and our experiments reveal that DL-MAC exhibits superior performance over traditional algorithms in both single and multi-channel environments and also outperforms single-function approaches in terms of overall performance. Additionally, the performance of DL-MAC remains robust, unaffected by channel switch overhead within the evaluated range.

\end{abstract}

\begin{keyword}

 Deep learning\sep Channel access\sep Rate adaptation\sep Channel switch



\end{keyword}

\end{frontmatter}


\section{Introduction}
\label{sect:Introduction}
Today, the demand for mobile smart devices and diverse internet applications is rising, leading to an expected increase in mobile network traffic to 403 exabytes per month by 2029 \cite{ericssonMobileDataTraffic2022}. With this trend, the number of devices employing short-range communication technologies such as Wi-Fi, Bluetooth, Zigbee, MulteFire, LTE-LAA, and NR-U is rapidly increasing, which results in a coexistence of dense devices in the unlicensed band, posing challenges to user experiences. This shift has intensified the heterogeneity of network environments, making the coordination between different protocols even more complex. 

In such a congested and heterogeneous wireless network, traditional medium access control (MAC) protocols are facing severe challenges. These MAC protocols were originally designed for homogeneous networks where all nodes follow the same rules. However, in complex and heterogeneous networks, they are inefficient at coordinating conflicts and interference among devices using different protocols, thereby struggling to meet the transmission efficiency and reliability requirements. For example, as observed in works \cite{jiang2017performance,yu2020non,guo2021ai}, the collision avoidance carrier sense multiple access (CSMA/CA) MAC protocol  in Wi-Fi networks, does not coexist effectively with other MAC protocols, including even the simplest ones like time division multiple access (TDMA) MAC protocol.

\begin{figure}[!t]
	\centering
	\includegraphics[width=2.5in]{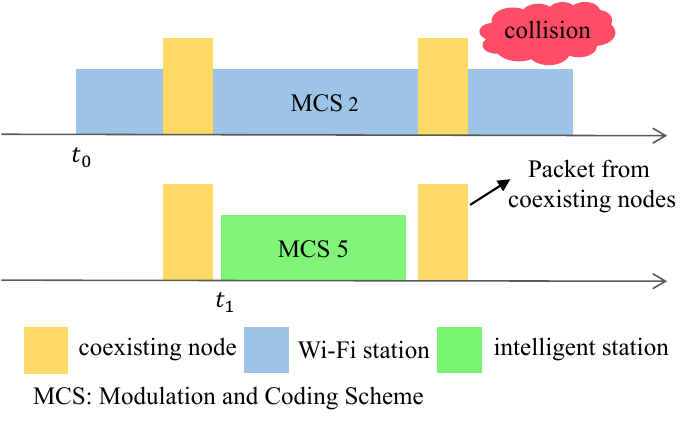}
	\caption{A toy example of traditional and intelligent protocol transmission.}
	\label{fig:toy_example}
\end{figure}

Another issue that deserves attention is that in traditional network protocol stacks, distinct operational functions are often designed as independent modules. The algorithm designed within each module focuses solely on a specific function and lacks collaborative operation with other modules. Take rate adaptation at the physical layer and channel access at the MAC layer as examples. Rate adaptation algorithms typically focus on adapting to the path loss of wireless channels, whereas channel access mechanism (e.g., CSMA/CA in Wi-Fi) is primarily concerned with collision avoidance. By shifting the paradigm and developing a novel MAC algorithm in a holistic manner that joints channel access and rate adaptation, it can potentially achieve substantial performance improvements.

Fig. ~\ref{fig:toy_example} illustrates a toy example of a heterogeneous coexisting network, showing the behavioral differences between traditional and intelligent protocols during transmission. As shown in Fig. ~\ref{fig:toy_example}, the blue and green rectangles depict a transmitting node of interest under two different protocols, while the orange rectangle represents concurrent transmission from a coexisting node. In the upper diagram, a Wi-Fi station (STA) utilizes the traditional CSMA/CA mechanism for channel access, employing a commonly adopted rate adaptation algorithm such as Intel I\textsc{wl} \cite{grunblatt2019simulation} to dynamically adjust its transmission rate. At time $t_0$, after the requisite carrier sensing and backoff, the Wi-Fi STA decides to transmit using a modulation and coding scheme (MCS\footnote{Herein, it is noted that rate adaptation and MCS selection both refer to the task of adjusting transmission rates based on the quality of the wireless channel, and we use these terms interchangeably throughout this paper.}) indexed by 2, as determined by the rate adaptation algorithm. We can see that a collision occurs between the Wi-Fi STA and the coexisting node, due to the Wi-Fi STA$'$s inability to predict the transmission behavior of the coexisting node. In contrast, the lower diagram showcases an intelligent STA that, at time $t_1$, leverages its learned insights of the wireless environment to predict the activity of coexisting nodes. This prediction capability allows it to make a precise channel access decision, coupled with an appropriate MCS selection—in this case, MCS 5—which avoids collision and enables a higher transmission rate. As depicted in Fig. ~\ref{fig:toy_example}, the key to designing an intelligent MAC algorithm lies in the ability to discern the transmission behavior of coexisting nodes by observing the wireless environment. This is achieved by employing a state-of-the-art data-driven deep learning framework. Consequently, the MAC protocol is nothing but a neural network model that processes a temporal sequence of channel sensing data to determine joint channel access and rate adaptation strategies. The input data--the channel sensing data, are readily extracted from the energy detection module common in wireless devices, enabling the neural network to infer whether and at what rate to access the channel based on historical spectrum conditions.

Furthermore, the channel switch mechanism can effectively cope with the dynamic changes in channel quality within crowded and heterogeneous wireless networks. However, traditional network protocols often lack the necessary flexibility and the ability to respond rapidly to changes in network conditions when designing channel switch mechanisms. Traditional switching strategies tend to be based on static settings and are incapable of adjusting to real-time changes in network conditions. Such static methods are less effective when the network environment changes rapidly, especially when channels become congested or interference intensifies. Moreover, traditional channel switch algorithms often overlook the communication needs of users and quality of service parameters, which may impose unnecessary delays on sensitive traffic under high network loads. Additionally, strategies involving frequent channel switches lead to intermittent link disruptions and exacerbate performance degradation due to poor decision-making. In this context, a more flexible and dynamic channel switch mechanism would be more suited to the needs of this complex and heterogeneous networks.

In response to the challenges of traditional protocols, this paper presents a deep-learning-based MAC paradigm, dubbed DL-MAC, which offers a holistic approach that integrates channel access, rate adaptation and channel switch into a unified system. Specifically, the DL-MAC protocol is a deep neural networks-based paradigm that takes spectrum sensing sequences freely acquired from wireless devices as input. By utilizing the advanced feature extraction and pattern recognition capabilities of deep learning, DL-MAC can extract information from these spectrum sensing sequences, enabling more efficient cross-layer decision-making on channel selection and the joint design of channel access and rate adaptation operating on a specific channel. To the best of our knowledge, we are the first to propose an integrating of channel access, rate adaptation and channel switch, and we have validated its performance on a real-world dataset, which has been made publicly available on GitHub for academic and research use.\footnote{The dataset is available on https://github.com/postman511/Wireless-Signal-Strength-on-2.4GHz-WSS24-dataset.}

Overall, the contributions of this work are summarized as follows:
\begin{itemize}
	\item We propose a deep-learning-based protocol, dubbed DL-MAC. This protocol integrates multiple functionalities, including channel access, rate adaptation and channel switch, aiming to optimize the spectrum efficiency and enhance the adaptability of wireless networks. 
	\item We have sampled wireless spectrum data from the 2.4 GHz industrial, scientific, and medical (ISM) frequency band in the real world and have made this dataset open-source and available to the GitHub community for public use. We have further verified the performance of the DL-MAC protocol using this real-world dataset. The results show that DL-MAC has superior performance of throughput and lower latency compared to traditional algorithms in both single-channel and multi-channel environments.
	\item To evaluate the performance gains yielded by the joint design of channel access and rate adaptation, we compared it with single-feature designs. These single-feature designs involve either a DL-based channel access combined with traditional MCS selection, or traditional channel access paired with a DL-based rate adaptation strategy. The experimental results show that the joint design of DL-MAC, when operating on a single channel, has significantly improved performance in terms of throughput and mean delay compared to designs based on a single feature.
	\item To confirm the protocol efficiency of DL-MAC in multi-channel settings, we assessed the potential performance overhead incurred by channel switch. The experiments demonstrate that, the performance of the DL-MAC protocol is not significantly affected by channel switch overhead within the tested range, proving its effectiveness in multi-channel settings.
\end{itemize}

\subsection{Related works}
In the IEEE 802.11 standard, CSMA/CA serves as the core mechanism for channel access. Early research on the CSMA/CA protocol mainly focused on optimizing parameters and improving algorithms, such as adjusting the size of the contention window  \cite{ksentini2005determinist,deng2008contention,zhou2015adaptive}, refining the backoff algorithm  \cite{li2009new,almotairi2013inverse,shurman2014n}, and enhancing the efficiency of the request to send/clear to send (RTS/CTS) mechanism  \cite{kong2004adaptive,mathew2015performance}. With the advancement of artificial intelligence (AI) technology, learning-based optimization strategies have also been proposed for the optimization of these parameters and enhancement of algorithms  \cite{amuru2015send,abyaneh2019intelligent,wydmanski2021contention,guo2022multi}. Furthermore, the work  \cite{pasandi2020mac} introduced a new method for designing network protocols based on DRL, where DRL agents can selectively combine different functionalities to enhance the performance of the protocol. However, this approach is only targeted at the CSMA protocol and has yet to provide a solution for the issues between different MAC protocols in heterogeneous networks. Note that, all these improvements to traditional algorithms, as well as learning-based methods, primarily focus on enhancing node throughput and access fairness, they do not sufficiently address the issue of coexistence among nodes using different protocols in complex and heterogeneous networks. In addressing the challenges of heterogeneous networks, the authors in \cite{yu2020non} and  \cite{yu2019deep} utilized DRL to design multi-access techniques tailored for heterogeneous networks, aiming to maximize throughput and ensure fairness in access while coexisting with other distinct protocols. However, these studies being based on simulated data, have not validated the performance of the proposed protocol in realistic scenarios and lack consideration in terms of rate adaptation.

Beyond the MAC layer$'$s channel access mechanisms, rate adaptation algorithms (RAA) at the physical layer have also been extensively studied. The ARF \cite{kamerman1997wavelan} algorithm dynamically adjusts the data transmission rate based on consecutive transmission successes, increasing or decreasing the rate accordingly, while SampleRate  \cite{bicket2005bit} and Minstrel  \cite{arif2017evaluation} schemes optimize rate selection using different methods based on past performance and statistics. Additionally, the Intel I\textsc{wl}  \cite{grunblatt2019simulation} algorithm also dynamically adjusts rates, but it specifically uses historical throughput and transmission count data to inform its adjustments. However, these algorithms aim to predict and adjust to the optimal future transmission rate based on past network performance data, employing certain exploration strategies to find the suboptimal rate. Recently, several studies have adopted AI technologies to achieve more accurate predictions and better adaptability in response to changing network conditions. The authors in \cite{li2020practical} used a three-layer fully connected network model to predict the best rate and adjust it according to real-time network conditions. Another work \cite{chen2021experience} utilized deep Q-learning to adapt to current link quality and channel conditions, implementing an intelligent rate adaptation algorithm. Apart from the above, the authors in  \cite{zhang2019deep} proposed a deep reinforcement learning-based MCS selection algorithm in heterogeneous networks. However, all these algorithms primarily solely focus on enhancing the performance of their specific function of rate adaptation, without taking into account other functionality, such as channel access.

What$'$s more, dynamic channel switch is also a key research topic for enhancing wireless network performance, especially in high-density and dynamically changing environments. Early studies mainly focused on static channel allocation strategies, which often rely on fixed attributes of channels, such as signal strength or interference levels  \cite{haidar2008channel,monteiro2012optimal}. However, these methods perform poorly under dynamically changing network conditions. As demands for network performance have increased, researchers have begun to shift towards dynamic channel selection that considers real-time network status. Representative works in this area include adaptive methods based on channel occupancy \cite{da2009tdcs}, network congestion \cite{musaddiq2014distributed}, and channel interference \cite{chakraborty2016dynamic}. The introduction of AI techniques has opened up new possibilities for dynamic channel selection. The authors in \cite{jeunen2018machine} employed machine learning to identify bad neighbors in Wi-Fi networks and allocate the optimal channel for each network to maximize the overall network performance. In another study  \cite{lee2023deep}, the authors proposed a deep-learning-based channel allocation scheme aimed at minimizing co-channel interference among APs, by utilizing a deep neural network to predict the channel allocation of each AP to minimize CCI. Although data-driven channel selection strategies can optimize channel selection decisions based on real-time network conditions and historical data, current research results mainly rely on outcomes obtained in simulation environments and have not been widely validated in real-world environments to prove their performance. This means that while these strategies perform well in theory and simulated tests, their actual effectiveness and reliability under the complex and constantly changing conditions of the real world remain to be further explored and confirmed.

In our previous work \cite{xin2022deep}, we developed DL-MAC protocol that focused on the joint design of channel access and rate adaptation on a specific channel to maximize spectrum efficiency. Differing from \cite{xin2022deep}, this paper extends DL-MAC protocol to operate channel access from one channel to multiple channels, i.e., incorporating the DL-MAC protocol with the ability of channel switch. In addition, we evaluated the effectiveness of the protocol in terms of throughput, but did not validate its performance in terms of packet transmission delay. In this work, DL-MAC is designed to enhance AI devices to operate in a multi-channel environment, by integrating channel switch functionality to adaptively select the most suitable channel for transmission.  Moreover, we have validated the proposed protocol, with a focus on its performance in terms of both throughput and mean delay.

\subsection{Outline}
The remainder of this paper is organized as follows: Section \ref{sect:background} presents the system scenario and preliminaries. Section \ref{sect:data_processing} describes the raw dataset and its processing, followed by Section \ref{sect:DL-MAC}, which details the specific implementation methods of the DL-MAC protocol. Numerical experiments conducted to verify the proposed DL-MAC protocol are discussed in Section \ref{sect: experimental results}. Finally, in Section \ref{sect:conclusion}, we conclude our work and discuss some possible research directions in the future.

\section{Background and preliminary} \label{sect:background}
In this section, we will provide background information on the system scenario and preliminaries in Wi-Fi networks.

\subsection{System scenario}
\begin{figure}[!t]
	\centering
	\includegraphics[width=2.5in]{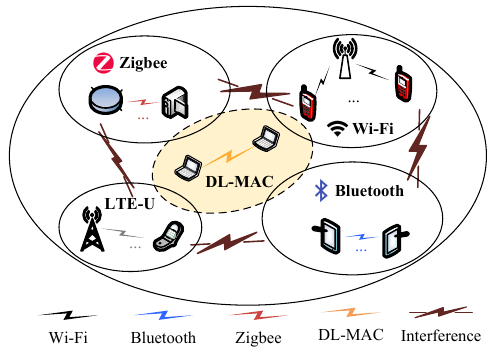}
	\caption{Coexistence of devices utilizing various protocols within the 2.4 GHz frequency band.}
	\label{fig:system_scenario}
\end{figure}

We consider a Wi-Fi network operating on the 2.4 GHz ISM frequency band. This network comprises both legacy Wi-Fi devices and AI devices, adhering to the CSMA/CA MAC protocol and the proposed DL-MAC protocol, respectively. In addition to Wi-Fi systems, devices operating under disparate wireless protocols such as Bluetooth, Zigbee, and LTE-U commonly coexist within the 2.4 GHz frequency band, as illustrated in Fig. ~\ref{fig:system_scenario}. These devices operate independently, each abiding by their respective protocols. Consequently, the concurrent operation of these diverse networks precipitates mutual interference, resulting in considerable spectrum inefficiency and deterioration of Wi-Fi communication quality.

\subsection{CSMA/CA mechanism}
The listen-before-talk (LBT) principle is fundamental to the CSMA/CA protocol, aimed at reducing packet transmission conflicts in shared channels. Under LBT, devices must sense that the channel is idle before initiating packet transmission. Additionally, the protocol incorporates a random backoff mechanism that introduces delays before transmissions, further reducing the chances of collisions. A device can transmit only after it has consistently sensed that the channel has been idle for the duration of a distributed coordination function interframe spacing (DIFS). After this DIFS period, the device enters a random backoff period, selecting a backoff counter and waiting for that duration, while pausing the backoff counter if the channel becomes busy. Once the backoff counter reaches zero, the device gains channel access for packet transmission. The CSMA/CA protocol employs a binary exponential backoff algorithm, where the backoff counter is uniformly chosen from (0, \textit{CW}-1), with \textit{CW} representing the contention window. Upon initial transmission failure, \textit{CW} begins at $CW_{min}$ and doubles up to $CW_{max}$, calculated as $2^PCW_{min}$, with $P$ being the maximum backoff exponent to limit transmission attempts.

In dense network environments, the CSMA/CA protocol is more prone to transmission  collisions, leading to extended backoff times for each device. The timing of transmissions is heavily influenced by the outcome (success or not) of the previous ones. This inspired us to design a specialized channel access protocol for independent decision-making regarding channel access.

\subsection{Rate adaptation}
In licensed bands, prevalent rate adaptation algorithms (RAA) rely on the signal-to-noise ratio (SNR). That is, the transmitter selects the MCS based on the SNR feedback or estimates from the inverse link. However, SNR-based methods often underperform in the high-interference unlicensed band because the optimal MCS depends not only on noise but also on interference from coexisting nodes. In practice, most RAA used in the unlicensed band employ sampling-based approaches. These algorithms sample statistical data, such as packet error rates (PER) and throughput across different rates, to evaluate channel conditions. Subsequently, they select the rate that has historically performed the best and employ it as the current optimal transmission rate. However, this approach can sacrifice some performance since it occasionally explores suboptimal rates to gather enough statistics. This motivates us to implement a more efficient RAA that directly selects rate options based on various channel conditions, without the necessity of exploring suboptimal rates.

\subsection{Channel switch}
Channel switch dynamically adapts to changing channel conditions, aiming to reduce interference and enhance network performance. Typically, an access point (AP) within a basic service set scans for and assesses the channel quality and congestion level of the surrounding channels. Based on this assessment, it chooses the optimal channel for communication with stations. However, in complex and dense networks, frequent channel switches due to signal noise and congestion can consume substantial time and resources, impacting user experience and network stability. Often, these switches are based on immediate signal quality assessments, which may overlook future channel uncertainties. Such a short-term focused approach can lead to decision biases and errors. Faced with this challenge, inspired us to develop an intelligent channel switch strategy. This approach effectively predicts future channel conditions under varying network conditions, thereby reducing frequent switching and avoiding biases and errors arising from short-term decision-making.

\section{Dataset and processing} \label{sect:data_processing}
Our proposed DL-MAC utilizes real-world wireless spectrum data collected from the 2.4 GHz ISM frequency band. This section briefly introduces the dataset and details the data processing procedure.

\subsection{Raw data collection}
The signal-to-interference-plus-noise ratio (SINR) is crucial for assessing wireless channel quality, influencing channel access and MCS selection. We propose using the received signal strength indicator (RSSI) as a proxy for SINR, exploiting its easy acquisition from energy detection modules in wireless devices for channel sensing. This approach simplifies the process of assessing wireless channel quality, leveraging RSSI$'$s availability and ease of measurement.

For real-world RSSI data, we measured values in the 2.4 GHz ISM band using the Ellisys Bluetooth Vanguard (BV1), capable of capturing radio frequency spectrum data across 79 channels, each with 1 MHz bandwidth, within the 2.4 GHz band. This includes signals from Bluetooth and other devices such as Wi-Fi, LTE-U, Zigbee, ANT, and more. The BV1$'$s real-time data collection from various networks offers comprehensive insights into the spectrum usage in this band.

We define each 1MHz sub-band as $f_n=2402+n$ (MHz), where $n=0,1,\cdots, N$, and the maximum value of $N$ is 78. RSSI values are sampled from each sub-band at a consistent interval $T_s$, set to 100$\mu$s. The sampling period comprises $L$ RSSI samples. For $N$ sub-bands, we periodically compile these RSSI samples into a comprehensive matrix as shown below:
\begin{small}
\begin{equation*}
	\setlength{\arraycolsep}{1.0pt}
	\boldsymbol{RSSI}=\left[\begin{array}{ccc}
	RSSI_{t_{1}, f_{0}}, & \cdots, & RSSI_{t_{1}, f_{N}} \\
	\vdots & \vdots & \vdots \\
	RSSI_{t_{L}, f_{0}}, & \cdots, & RSSI_{t_{L}, f_{N}}
	\end{array}\right].
\end{equation*}
\end{small}

In this matrix, each row represents the sampling data across different sub-bands at a single point in time, while each column captures the sampling data over time for the same sub-band. Thereby,  $RSSI_{t_{l},f_{n}}$, represents the $(l,n)$-th entry, where $t_{l}=t_{0}+T_{s}l$. Here, $t_0$ is the starting time, and $l$ represents the $l$-th sample, ranging from $1$ to $L$. This raw RSSI matrix encapsulates the entire data set, representing the spectrum of RSSI values across all the observed sub-bands.

Note that $L$ denotes the size of the sample set collected in our dataset. Due to the fine sampling interval of 100$\mu s$, the dataset is significantly large. Without loss of generality, in subsequent sections, we focus our design and experiments on data from the first two minutes of this extensive dataset to validate the performance of our proposed DL-MAC protocol. It is noteworthy that the raw dataset is open-source and available for public use.

\subsection{Data preprocessing}
In Wi-Fi networks, the mini-slot, $T_{slot}$, is the basic time unit, set at 9$\mu$s \cite{ieee2007ieee}. Thus, we process the raw RSSI data from the 79 sub-bands to align with Wi-Fi standards and convert it to fit the 13 Wi-Fi network channels. 

Initially, we align the raw RSSI data with Wi-Fi network time slots in the time domain. The sampling interval of RSSI data is ${T_s} = 100\mu$s, differing from the mini-slot, $T_{slot}$, in Wi-Fi networks. To align the time units of the CSMA/CA protocol with the RSSI data, we use linear interpolation for up-sampling with a factor of $T_s/T_{slot} \approx 11$. Specifically, for an interpolation point $t^\prime$, we identify the two closest time points $t_1$ and $t_2$ around it and use the original values at these points for linear interpolation. Assume that $RSSI_{i,f_j}$ represents the element at the $i$-th row and $j$-th column in the original matrix. Then, for the interpolation point $t^prime$ on sub-band $f_j$, the interpolated value $RSSI_{t\prime,f_j}$ can be calculated using the following formula:
\begin{small}
\begin{equation*}
\begin{split}
RSSI_{t^{\prime}, f_{j}}^{\prime}=&\frac{RSSI_{t_2,f_j}\!-\!RSSI_{t_1, f_j}}{t_2\!-\!t_1}\left(t^{\prime}-t_1\right)\!+\!RSSI_{t_1, f_j},
\end{split}
\end{equation*}
\end{small}where $t_1$ and $t_2$ are the nearest time points to $t^\prime$, located immediately to the left and right, respectively. 

After time domain processing, we map the data to Wi-Fi network channels. Wi-Fi networks typically have 13 overlapping channels, with the central frequency of the $i$-th channel defined as $F_i=2412+5(i-1)$ (MHz), for $i$ ranging from 1 to 13. We use average interpolation with a down-sampling factor of 21, corresponding to the bandwidth of a Wi-Fi channel. For instance, channel 6, covering 21 sub-bands from 2.427 GHz to 2.447 GHz $(f_{25},\cdots,f_{45})$, averages these sub-bands$'$ RSSI data for its value. Mathematically, the RSSI for the $i$-th Wi-Fi channel at time $t_l^\prime$ is ${RSSI}_{t_l^\prime, i}^\prime=\frac{1}{21}\sum_{f=F_i-10}^{f=F_i+10}{RSSI}_{t_l^\prime, f}^\prime$. Consequently, the final RSSI matrix is then obtained as follows:
\begin{small}
\begin{equation*}
	\setlength{\arraycolsep}{1.0pt}
	\boldsymbol{RSSI}^{\prime}=\left[\begin{array}{ccc}
	RSSI_{t_1^\prime, 1}^{\prime}, & \cdots, & RSSI_{t_1^\prime, 13}^{\prime} \\
	\vdots & \vdots & \vdots \\
	RSSI_{t_{L{^\prime}}^{\prime}, 1}^{\prime}, & \cdots, & RSSI_{t_{L{^\prime}}^{\prime}, 13}^{\prime}
	\end{array}\right].
	\label{eq_RSSI_matrix}
\end{equation*}
\end{small}

Note that, the original RSSI matrix dimensions are $L\times N$, where $L$ is the number of samples, and N represents the number of sub-bands at each sampling time point. Through the interpolation process in the time domain, the number of samples increases to $L^\prime=L \times \lfloor {T_s}/T_{slot} \rfloor$. Additionally, interpolation in the frequency domain modifies the number of columns in the original RSSI matrix. Initially, the matrix had $N$ columns corresponding to the original sub-bands; after interpolation, this scales to 13 columns, each representing the RSSI values for one of the 13 channels. Consequently, the dimensions of the interpolated RSSI matrix become $L^\prime \times 13$. Generally, if we focus on a particular channel, such as channel $i$, the RSSI matrix can be reduced to a vector, denoted as $\boldsymbol{RSSI}^{\prime}=[{RSSI}_{t_1^\prime, i}^\prime,\cdots,RSSI_{t_1^\prime, i}^\prime]^T$. For clarity in subsequent discussions, we will uniformly refer to the processed data as RSSI in the following. 

\begin{figure}[!t]
	\centering
	\includegraphics[width=2.5in]{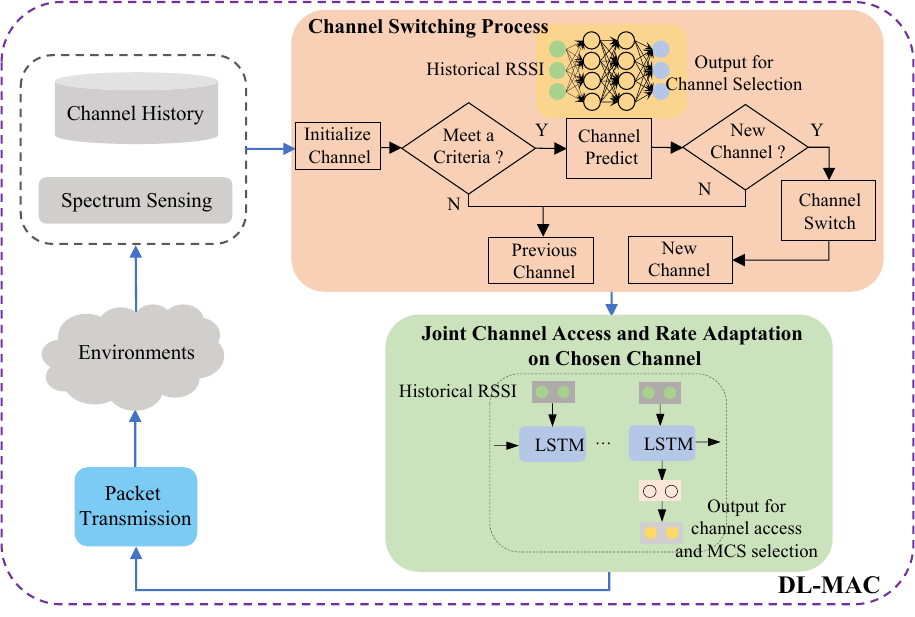}
	\caption{Overview of DL-MAC framework.}
	\label{fig:DL-MAC}
\end{figure}

\section{DL-MAC protocol} \label{sect:DL-MAC}
In this section, we will first provide an overview of our DL-MAC framework design,  then detail the joint channel access and rate adaptation mechanism for a specific channel, and finally, describe our approach for implementing channel switch in DL-MAC. 

\subsection{DL-MAC framework}
Fig. \ref{fig:DL-MAC} details the DL-MAC framework, emphasizing its two main components: channel switching process and the joint design of channel access and rate adaptation on selected channel. Unlike traditional methods, DL-MAC relies on sequential channel sensing results and employs deep learning for decision-making. It integrates a deep neural network (DNN) for channel selection decisions and a recurrent neural network (RNN) for joint channel access and rate adaptation on a selected channel. A fundamental requirement for DL-MAC is its capability to continuously sense all available channels in real-time, with the data stored in a constantly updated channel history database. This database provides the input data for the DNN and RNN, which then make decisions based on the input. When channel switch conditions are met, the DNN activates and determines the optimal channel for subsequent transmission. If the chosen channel matches previous selections, DL-MAC remains on the current channel; otherwise, it initiates switching. After switching, the RNN uses historical data of the selected channel to perform the joint design of channel access and rate adaptation. It should be noted that, when the DL-MAC protocol is designated to operate on a specific channel, it functions for single-channel use, and solely focuses on the joint design of channel access and rate adaptation, as delineated in our earlier work \cite{xin2022deep}.

\subsection{Joint channel access and rate adaptation}\label{subsect:jcara}
In this subsection, we explore the mechanism of joint channel access and rate adaptation for a specific channel, beginning with the essential data labeling process. Next, we detail the structure of the neural network involved, followed by outlining the specific implementation steps to give a comprehensive view of the procedure.

\subsubsection{Data labeling procedure}\label{jcara_data_label}
In our study, channel access and MCS selection are performed using supervised learning, with dataset labeling being a crucial initial step.

\begin{figure}[!t]
	\centering
	\includegraphics[width=2.5in]{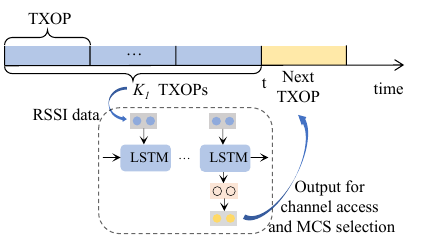}
	\caption{Illustration of data labeling and RNN.}
	\label{fig:RNN}
\end{figure}

We define a transmit opportunity (TXOP\footnote{A transmit opportunity (TXOP) is a designated period when a Wi-Fi STA can access the channel for data transmission. It sets the maximum time a STA can occupy the channel, defined here in terms of the number of mini-slots.}) spanning 120 mini-slots, and Fig. \ref{fig:RNN} illustrates the RNN alongside a training sample. At time $t$, we use historical RSSI data from the past $K_1$ TXOPs as input. The input data comprising $K_1$ sequences of 120 RSSI values each. The input$'$s label indicates the optimal channel access and MCS selection for the next TXOP (from $t+1$ to $t+120$), specifically the MCS from Table \ref{table1} that maximizes transmission rate without exceeding a target PER.

We define $MCS_i$ as the MCS indexed by $i$ in Table \ref{table1}. PER is dependent on SINR and the chosen $MCS_i$, expressed as $PER=f\left(SINR;MCS_i\right)$. With a constant SINR, PER increases monotonically with $MCS_i$. For each $MCS_i$, to maintain PER below a target $T_{PER}$, we calculate the minimum SINR ($SINR_i^{min}$) that meets $f\left(SINR_i^{min};MCS_i\right)=T_{PER}$. This allows us to establish the operational SINR range for each $MCS_i$ as $[SINR_i^{min},SINR_{i+1}^{min})$, where $MCS_i$ attains  maximum transmission rate within the target PER. Table \ref{table1} enumerates the minimum SINR requirements corresponding to various MCS indices for operational purposes \cite{MCStable}. We also outline a detailed workflow for the data labeling process as follows:
\begin{itemize}
	\item Calculate average RSSI: For each time $t$, we gather RSSI data $D_{K1}^t=\{RSSI_{t-120K_1+1},\cdots,RSSI_t\}$ from the previous $K_1$ TXOPs, which span $120K_1$ mini-slots. We then observe the RSSI values for the upcoming TXOP time intervals, i.e., $\{RSSI_{t+1},\cdots,RSSI_{t+120}\}$, and calculate their average RSSI, given by: $\overline{RSSI}=\frac{1}{120}\sum_{l=1}^{l=120}{RSSI_{t+l}}$.
	\item Calculate SINR: We calculate the average SINR at the receiver by applying the formula $SINR=P_r-\overline{RSSI}$. In this calculation, SINR is calculated as the difference in decibels between $P_r$ and $\overline{RSSI}$, where $P_r$ denotes the received signal power at the receiver. In practical systems, $P_r$ can be easily estimated at the transmitter. For our experiments, we assume $P_r$ to be -65dBm for simplicity, without loss of generality. 
	\item Lookup MCS: We determine the MCS for the upcoming TXOP based on the calculated SINR and the minimum SINR in Table \ref{table1}. For example, if SINR is 12dB, $D_{K1}^t$ is labeled as $MCS_3$.
\end{itemize}

\begin{table}[!t]
\centering
\caption{MCS indices and corresponding minimum SINR required.}
\resizebox{80mm}{20mm}{
\begin{tabular}{|c|c|c|c|c|}
\hline
\textbf{MCS Index} & \textbf{Modulation} & \textbf{\begin{tabular}[c]{@{}c@{}}Coding\end{tabular}} & \textbf{\begin{tabular}[c]{@{}c@{}}Rate(Mbps)\end{tabular}} & \textbf{\begin{tabular}[c]{@{}c@{}}Min. SINR(dB)\end{tabular}} \\ \hline
-1 & - & - & - & \textless 2 \\ \hline
0 & BPSK & 1/2 & 6.5 & 2 \\ \hline
1 & QPSK & 1/2 & 13 & 5 \\ \hline
2 & QPSK & 3/4 & 19.5 & 9 \\ \hline
3 & 16-QAM & 1/2 & 26 & 11 \\ \hline
4 & 16-QAM & 3/4 & 39 & 15 \\ \hline
5 & 64-QAM & 2/3 & 52 & 18 \\ \hline
6 & 64-QAM & 3/4 & 58.5 & 20 \\ \hline
7 & 64-QAM & 5/6 & 65 & 25 \\ \hline
8 & 256-QAM & 3/4 & 78 & 28 \\ \hline
\end{tabular}}
\label{table1}
\end{table}

\subsubsection{Neural network}
The RNN used for joint channel access and rate adaptation includes an input layer, two hidden layers, and an output layer. The input layer has a size of 120, corresponding to the RSSI data from the TXOP. The two hidden layers consist of a long short-term memory (LSTM) layer and a feedforward layer. The output layer generates a probability distribution over MCS indices using a SoftMax layer. This output enables the prediction of joint channel access and MCS selection at time $t$, supporting the DL-MAC$'$s operation on a specific channel.

\subsubsection{Implementation of joint channel access and rate adaptation}\label{subsec:imp_jcara}
After labeling, we train the RNN using the Adam algorithm \cite{kingma2014adam} and cross-entropy loss with the labeled RSSI data. The trained model is then used for joint channel access and rate adaptation.

Initially, the AI device starts with an idle state and listens to the channel. At each mini-slot $t$, it collects one RSSI data and then puts it into the tail of the RSSI data queue. Subsequently, it inputs the latest $120K_1$ RSSI data into the RNN to select an  $MCS_i$. In cases where $MCS_i$ equals $-1$ (indicating no channel access), the AI device continues listening to the channel and acquiring new RSSI data at the next mini-slot $t+1$. If the RNN output indicates transmission, the device transmits the earliest packet in its buffer using the selected $MCS_i$ (not equal to $-1$) in the upcoming 120 mini-slots. After transmission, it receives ACK or NACK feedback. During transmission, due to half-duplex constraints, the device cannot receive any signals from the wireless channel. Therefore, it becomes necessary to generate RSSI data in a handcrafted manner for this packet transmission time after the transmission is finished. This data, reflecting interference levels based on transmission success or failure, is added to the queue for subsequent RNN predictions.

When $MCS_i$ is selected for packet transmission. If packet is successfully transmitted, it is supposed that interference from other devices is low. Consequently, the RSSI data is generated uniformly within the range from the RSSI value corresponding to $SINR_i^{min}$ to the RSSI value corresponding to $SINR_8^{min}$. Conversely, if packet transmission fails, this suggests high interference from other devices. In this case, the RSSI data is uniformly generated within the range from the RSSI value corresponding to $SINR_{-1}^{min}$ to the RSSI value corresponding to $SINR_{i-1}^{min}$.

\subsection{Chanel switch}
The channel switch mechanism, leveraging SINR metrics, enables AI devices to assess all accessible channels and determine the most suitable one guided by a DNN model that predicts future network conditions. Upon choosing the optimal channel, AI devices perform joint channel access and rate adaptation for it, allowing adaptation to dynamic network conditions. This subsection outlines the data labeling process, neural networks, and workflow involved in channel switch.

\subsubsection{Data labeling process}
Similar to joint design of channel access and rate adaptation, channel switch is also executed within a supervised learning framework. Consequently, it is essential to label the RSSI data used for channel switch. Likewise, each TXOP consists of 120 mini-slots.

\begin{figure}[!t]
	\centering
	\includegraphics[width=2.5in]{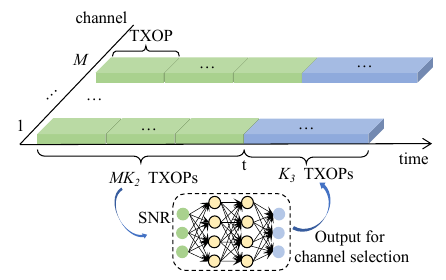}
	\caption{Illustration of data labeling and DNN.}
	\label{fig:DNN}
\end{figure}

Assuming AI devices can dynamically switch among the $M$ available channels, as shown in Fig. \ref{fig:DNN}. We use historical RSSI data from all $M$ available channels to predict the channel qualities in the near future, and select the channel with the highest predicted quality for channel access and rate adaptation for the next period, i.e., the duration time of K3 TXOPs. Specifically, for each channel within the $M$ channels, we collect RSSI data from the past $K_2$ TXOP intervals and observe RSSI data for the upcoming $K_3$ TXOP periods. For the $m$-th channel, such data is denoted as $D_{m,K2}^t=\{RSSI_{m, t-120K_2+1}, \cdots, RSSI_{m, t}\}$ for past intervals and $D_{m,K3}^t=\{RSSI_{m,t+1},\cdots,RSSI_{m,t+120K_3}\}$ for upcoming intervals, respectively. Subsequently, the average SINR values for these datasets, denoted as $SINR_{m,K_2}^t$ and $SINR_{m,K_3}^t$, are calculated in accordance with the methodology described in the data labeling process of Section \ref{subsect:jcara}. Consequently, the feature set for a sample at time $t$ is $SINR_{K_2}^t=\{SINR_{m,K_2}^t\}_{m=1}^M$, and its corresponding label is the channel from these $M$ options with  the highest average $SINR_{m,K_3}^t$ in the upcoming $K_3$ TXOP periods. This effectively identifies the optimal channel for the specified time intervals.

\subsubsection{Neural network}
The DNN designed for channel switch adopts a fully connected architecture consisting of five layers: an input layer, three hidden layers, and an output layer. The input layer is designed to handle a feature vector of dimensions $MK_2$, where $M$ represents the number of accessible channels and $K_2$ denotes the quantity of independent SINR values per channel. Following each hidden layer, a rectified linear unit (ReLU) activation function is employed. The network concludes with an output layer, which utilizes a SoftMax activation function to generate a probability distribution for each channel. 

\subsubsection{Implementation of channel switch}
After labeling process, we train our DNN model using the Adam stochastic gradient descent algorithm, coupled with cross-entropy loss function. Upon training completion, the model is  deployed onto AI devices.

Initially, the AI device sets a time interval $T_c$ for channel switch and selects a channel $ch_i$ at random for joint channel access and rate adaptation, as elaborated in Section \ref{subsec:imp_jcara}. Note that, the AI device maintains individual RSSI queues for each channel, appending collected RSSI data to respective queues at every mini-slot. Additionally, the AI device initiates a timer based on a preset interval. When this timer reaches the predetermined $T_c$ and the device is idle, the channel switch process is initiated; otherwise, it maintains its operations on the current channel. Upon entering the channel switch phase, which involves transitioning from one channel to another, the AI device undergoes a series of configurations and adjustments, including modulating its receiving and transmitting frequencies to match the new channel. During this process, the device is temporarily unable to receive signals from any channel. It resumes signal reception only after completing these adjustments and fully transitioning to the new channel. For simplicity, our primary focus is on the channel selection aspect. Other related configuration activities during the channel switch are collectively accounted for under the time $T_d$, known as the channel switch overhead.

The AI device selects the channel $ch_j$ with the highest probability as the newly selected channel. If $ch_j$ differs from the previously chosen channel $ch_i$, a switch occurs, followed by joint channel access and rate adaptation on $ch_j$. Otherwise, the device continues operating on the current channel $ch_i$. Due to the half-duplex constraints, the AI device cannot receive signals and RSSI compensation is necessary for all channels during transmission and switching phases. For the selected channel, RSSI compensation follows the procedures outlined in the implementation of joint channel access and rate adaptation. For other channels, compensation values are randomly chosen from the entire RSSI range are used.

\section{Experimental results} \label{sect: experimental results}
This section outlines experiments that assess the performance of DL-MAC protocol. We will detail the experimental setup, performance metrics, comparison algorithms, and finally presents the results.

\subsection{Experimental settings}
\subsubsection{Sampling environment}
\begin{figure}[!t]
	\centering
	\includegraphics[width=2.5in]{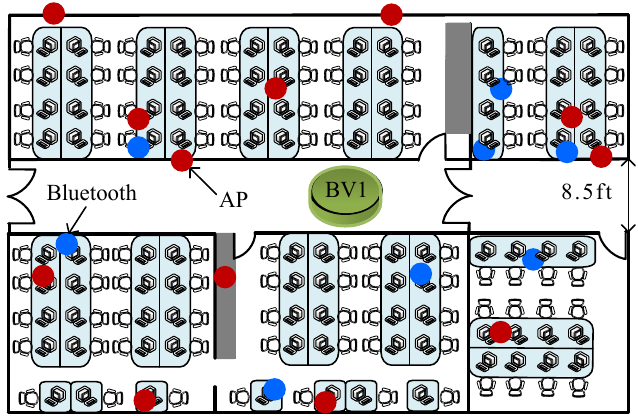}
	\caption{A schematic layout of wireless data sampling setup in the laboratory environment at Shenzhen University.}
	\label{fig:lab_layout}
\end{figure}

Our data sampling was conducted in a laboratory setting at Shenzhen University. In the lab, we established an environment with 12 Wi-Fi APs and 8 Bluetooth devices for data collection. The positions of these APs and Bluetooth devices were randomly assigned, with our data sampling hardware, BV1, strategically placed at the center. The layout of this wireless sampling setup is depicted in Fig. ~\ref{fig:lab_layout}.

\subsubsection{Platform and tools}
For the evaluation of DL-MAC and comparative benchmarks, we developed a simulator using Python. Our simulator ran on a computer configured with an Intel Core i9-10900K CPU and an NVIDIA GeForce RTX 3090 Founders Edition GPU. For neural network training purposes, PyTorch \cite{paszke2017automatic} was selected as our deep learning platform.
\subsubsection{System parameters} \label{sect:system parameters}
The fundamental time unit in our experiments is the mini-slot, set to a duration of 9$\mu$s, aligning with Wi-Fi network standards for compatibility. Regarding traffic generation, packet arrivals at the transmitting buffer follow a Poisson distribution, with the arrival rate $\lambda$ (indicating the average number of packets arriving per mini-slot) set to 0.18. Each packet is configured with a default payload size of 1500 bytes, and the buffer is designed to accommodate up to 10 packets. Additionally, we have defined three distinct values for TXOPs: $K1$, $K2$, and $K3$, set at 3, 5, and 2, respectively.

In the configuration of the CSMA/CA mechanism, the interference threshold is established at -75dBm. The DIFS is designated a duration of 36$\mu$s. Furthermore, the contention window parameters are defined with $CW_{min}$ set to 32 and $CW_{max}$ to 1024. 

\subsection{Performance metrics}
In this experiment, we assess the performance of DL-MAC, focusing on its ability to maximize throughput while minimizing packet delay. The following metrics are defined to quantify its performance:
\begin{itemize}
	\item Throughput: This is quantified as the number of successfully transmitted bits per mini-slot, providing a measure of the data transmission efficiency of the system.
	\item Packet Delay: Defined as the number of mini-slots that elapse from the moment a packet is generated until it is successfully transmitted, this metric measures the latency in packet delivery.
	\item Mean Delay: Representing the average delay experienced by all successfully transmitted packets, the mean delay provides an overall assessment of the system latency.
\end{itemize}

\begin{figure}[!t]
  \centering
  \begin{subfigure}{0.85\linewidth}
    \includegraphics[width=\linewidth]{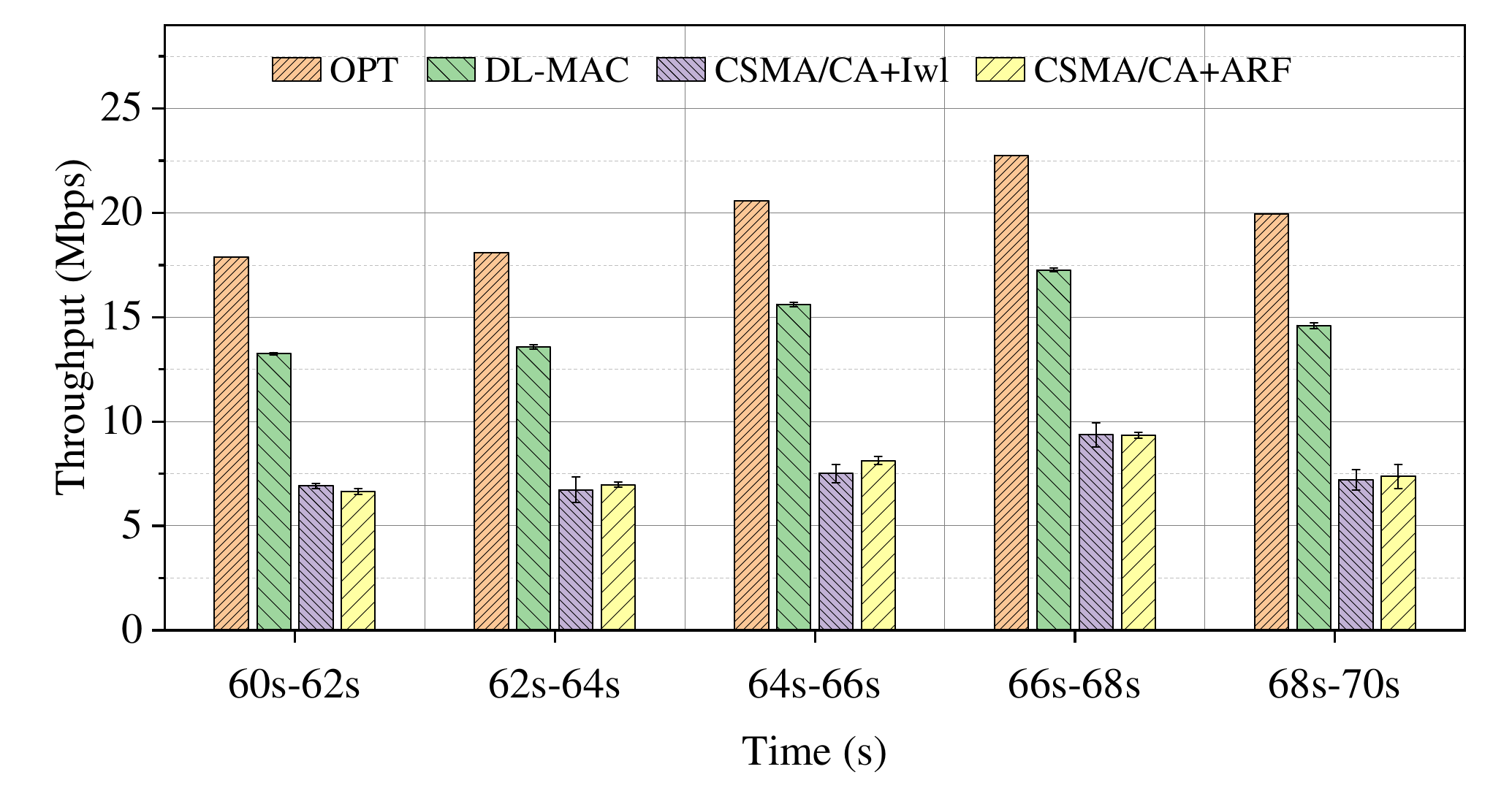}
    \subcaption{Throughput}
    \label{fig:ch6_performance-comparison_a}
  \end{subfigure}
  
  \vspace{2mm}
  
  \begin{subfigure}{0.85\linewidth}
    \includegraphics[width=\linewidth]{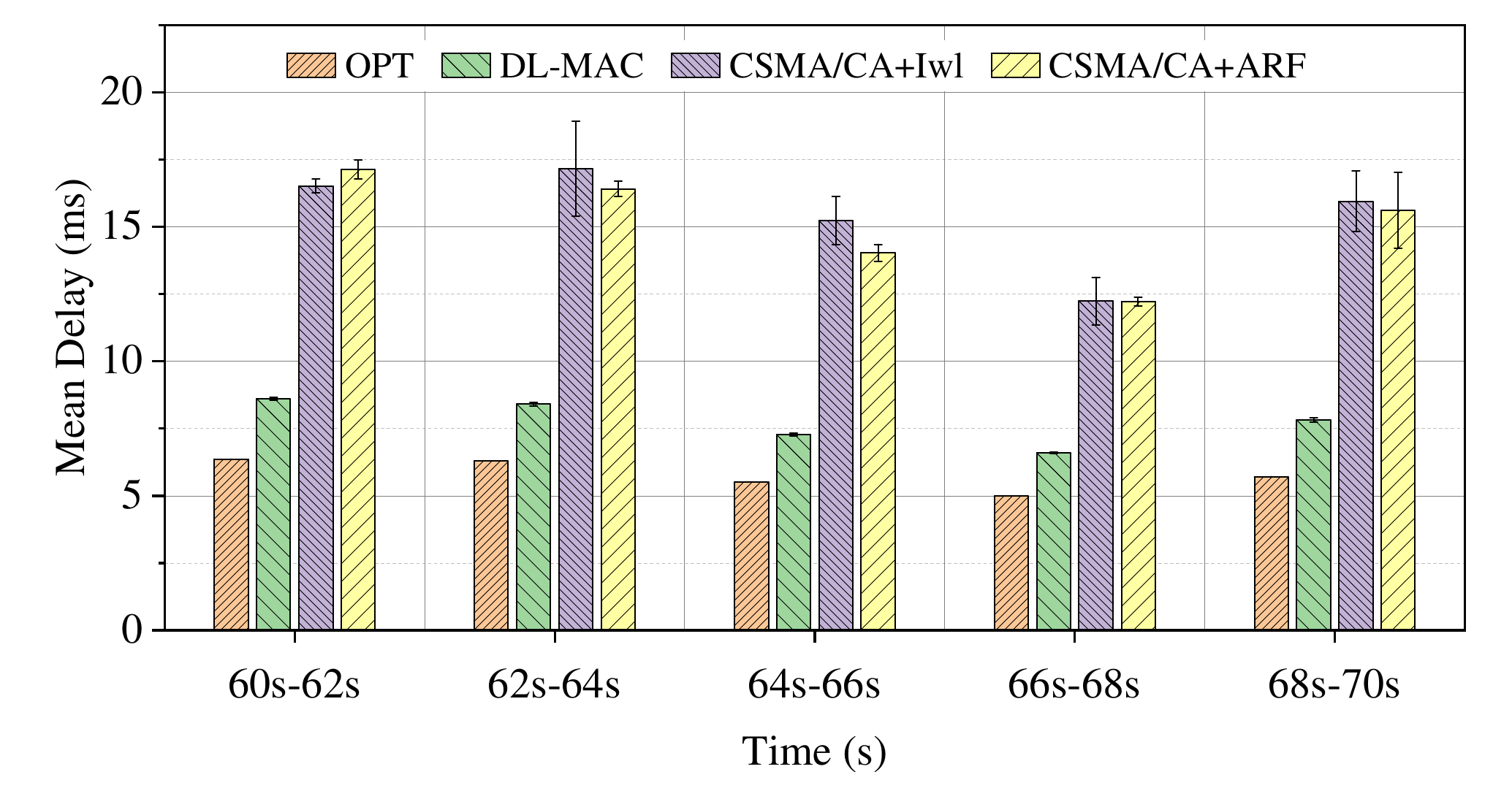}
    \subcaption{Mean delay}
    \label{fig:ch6_performance-comparison}
  \end{subfigure}
  
  \caption{Performance comparison of DL-MAC operating on channel 6: (a) Throughput; (b) Mean delay.}
  \label{fig:ch6_performance-comparison_b}
\end{figure}

\begin{figure*}[!t]
  \centering
  \begin{subfigure}{0.4\textwidth}
    \includegraphics[width=\linewidth]{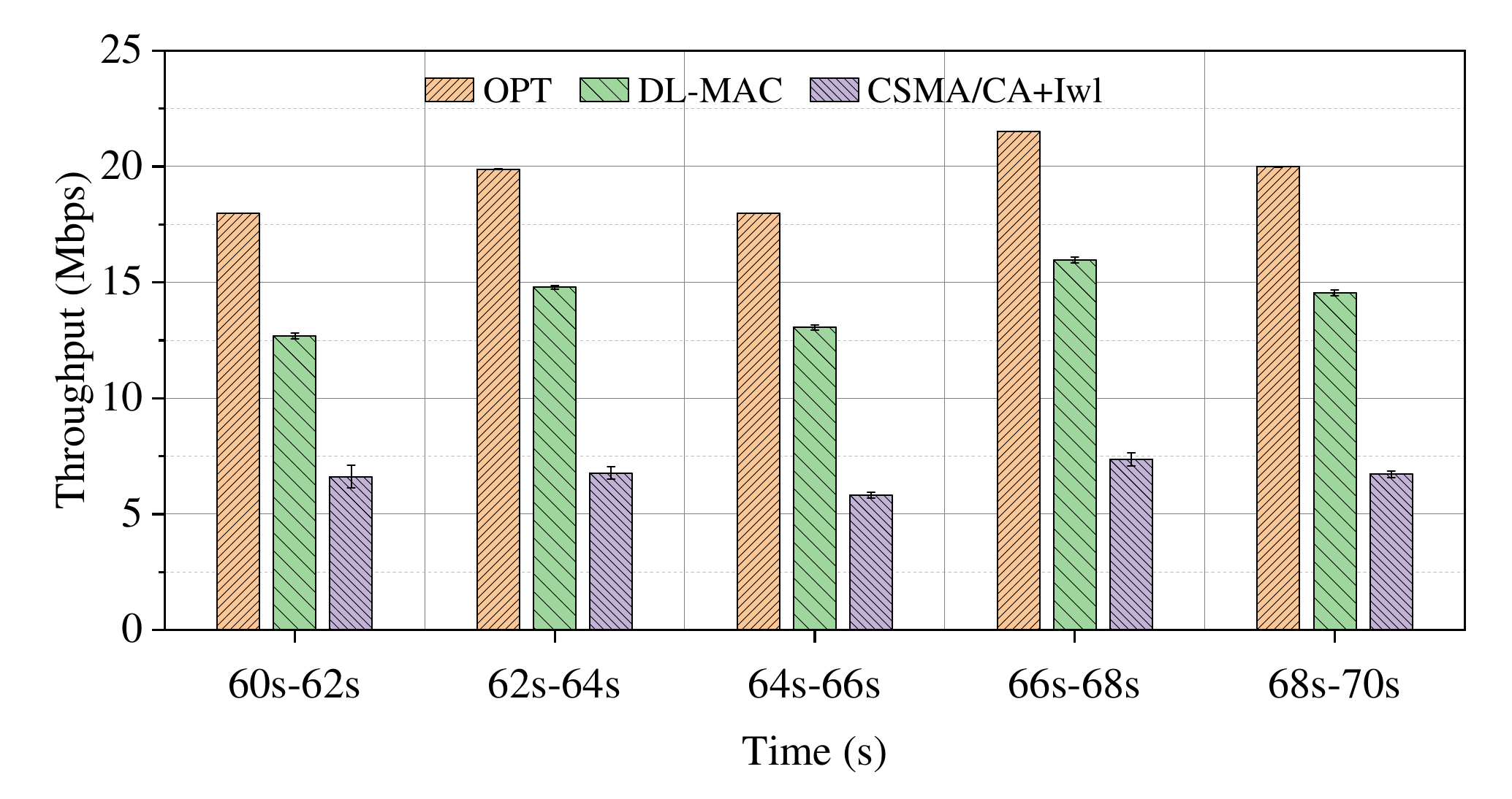}
    \subcaption{Throughput on channel 1}
    \label{fig:generalization-comparison_a}
  \end{subfigure}
  \begin{subfigure}{0.4\textwidth}
    \includegraphics[width=\linewidth]{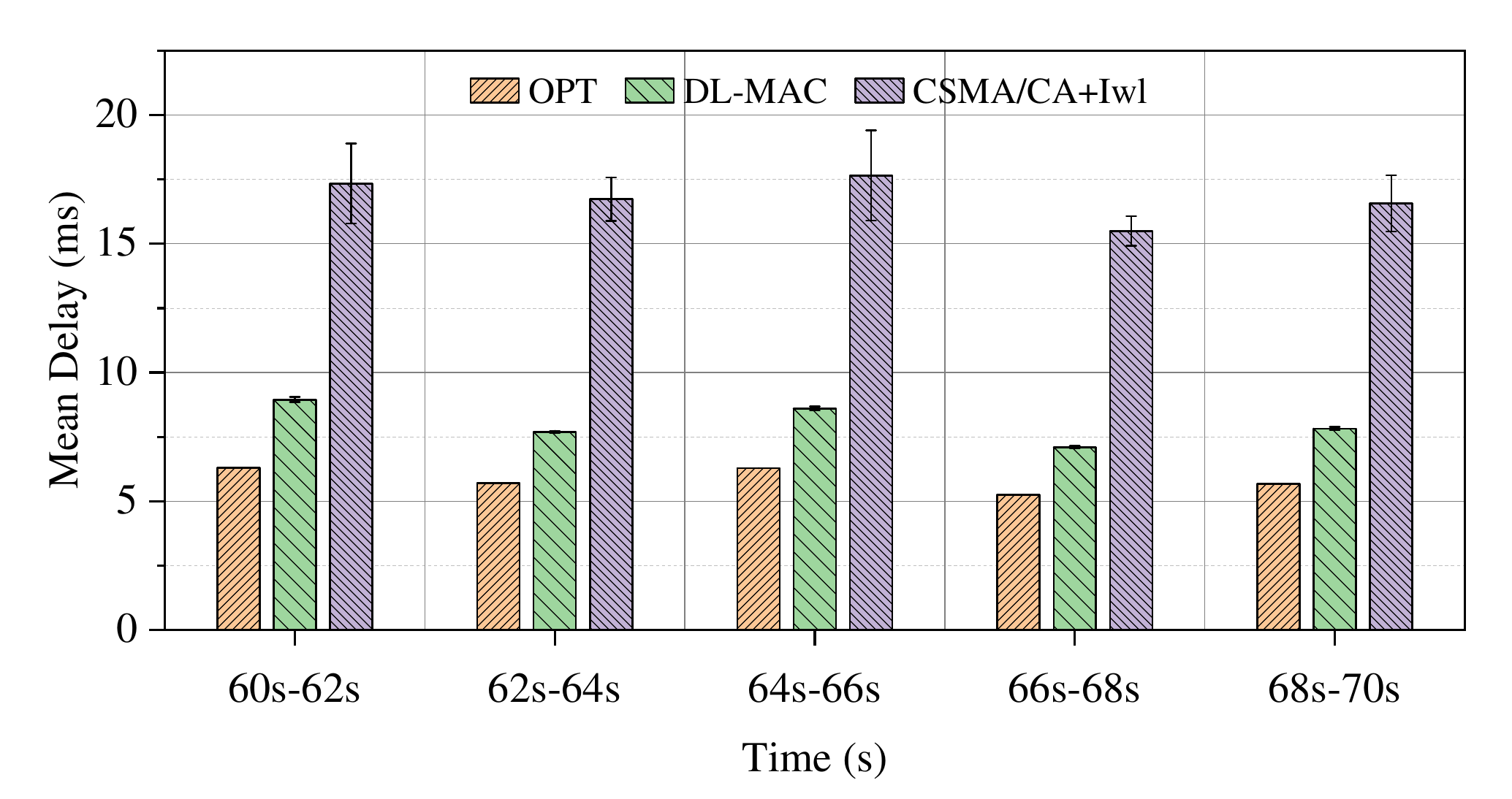}
    \subcaption{Mean delay on channel 1}
    \label{fig:generalization-comparison_b}
  \end{subfigure}
  
  \begin{subfigure}{0.4\textwidth}
    \includegraphics[width=\linewidth]{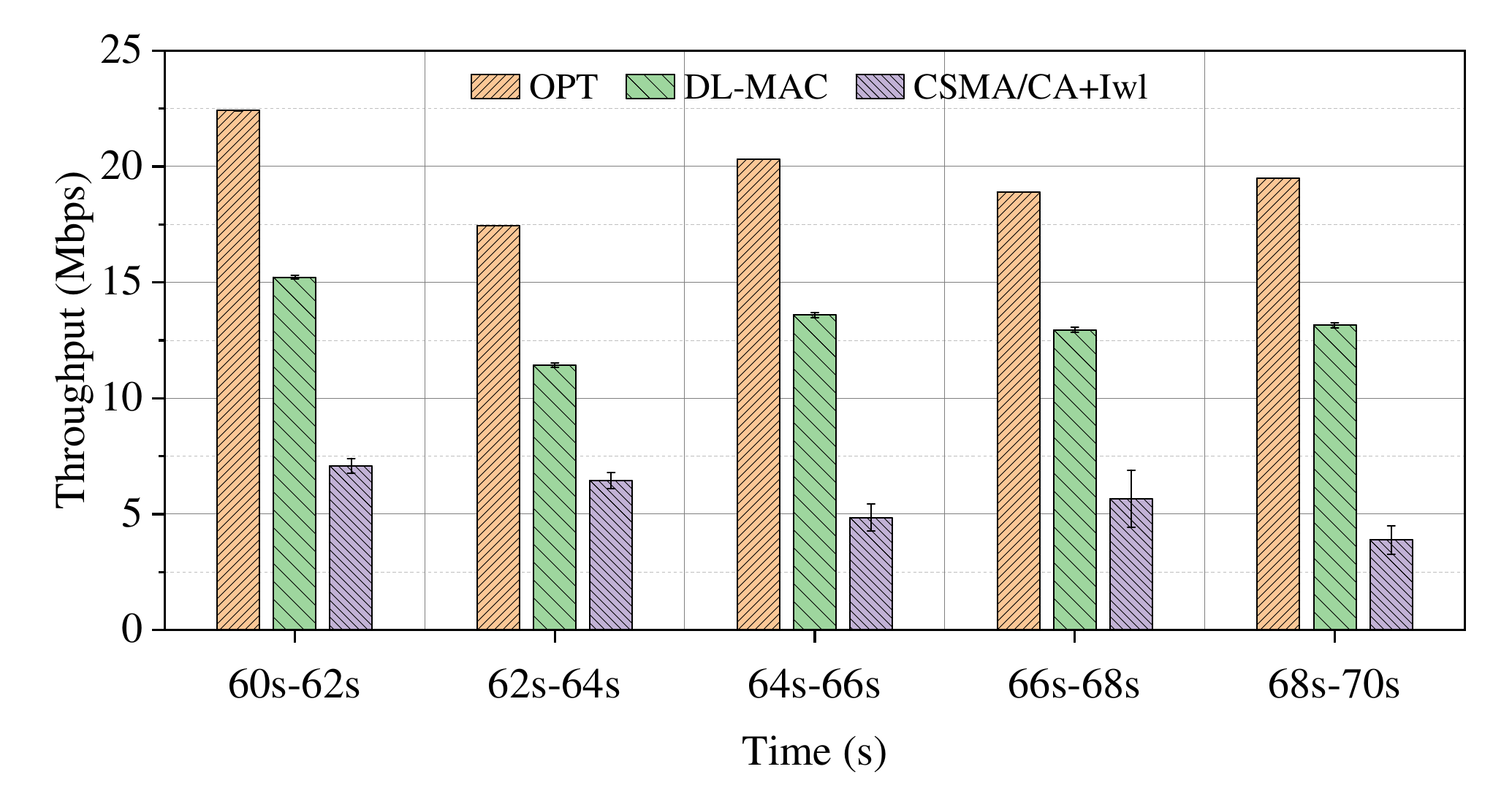}
    \subcaption{Throughput on channel 11}
    \label{fig:generalization-comparison_c}
  \end{subfigure}
  \begin{subfigure}{0.4\textwidth}
    \includegraphics[width=\linewidth]{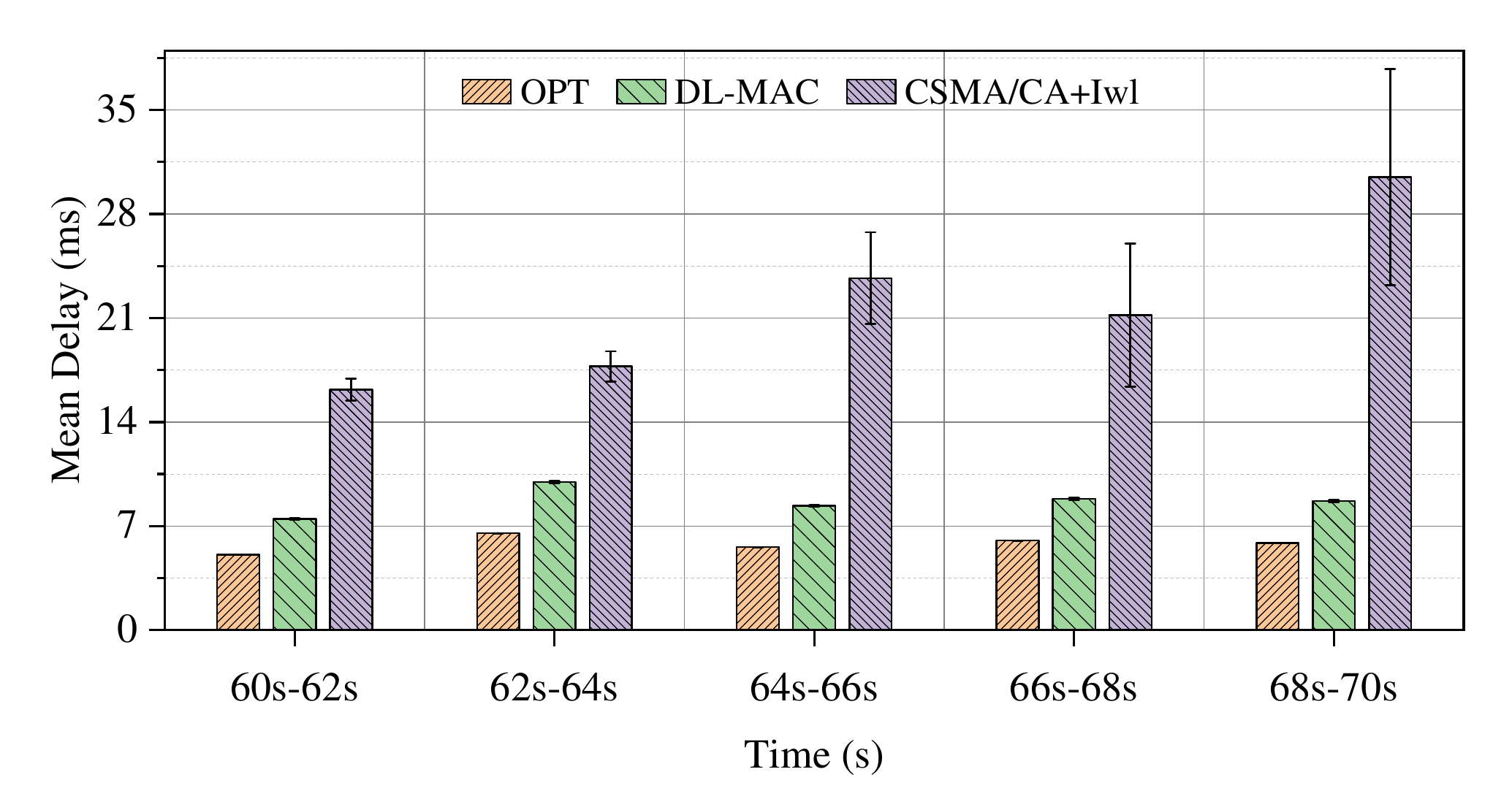}
    \subcaption{Mean delay on channel 11}
    \label{fig:generalization-comparison_d}
  \end{subfigure}
  
  \caption{Generalization performance comparison of DL-MAC on channels 1 and 11: (a) Throughput on channel 1; (b) Mean delay on channel 1; (c) Throughput on channel 11; (d) Mean delay on channel 11.}
  \label{fig:generalization-comparison}
\end{figure*}

\subsection{Algorithms for comparison}
To evaluate the performance of DL-MAC, we employ several comparative approaches. The optimal transmission policy, derived from the actual real-world sampled RSSI data, serves as the upper bound for MAC performance. This policy determines the optimum for joint channel access and MCS selection for a given channel. Additionally, we use the traditional CSMA/CA protocol and rate adaptation algorithms servers as a conventional benchmark. Moreover, to demonstrate the advantages of our joint design, we compare individual deep-learning-based methods for channel access and MCS selection against our joint design. The specific algorithms and combinations used in our benchmarking are as follows:
\begin{itemize}
	\item OPT: The optimal transmission scheme determines the best decisions for the joint design of channel access and MCS selection on a specific channel. We consider OPT as the upper bound for MAC performance operating on a single channel when DL-MAC operates on that channel. 
	\item CSMA/CA+I\textsc{wl}: This setup uses the CSMA/CA protocol for channel access, combined with the Intel I\textsc{wl} algorithm for MCS selection.
	\item CSMA/CA+ARF: In this configuration, CSMA/CA is used for channel access, while the ARF algorithm is employed for MCS selection. 
	\item CSMA/CA+DL/MCS: This approach pairs CSMA/CA for channel access with a deep-learning-based algorithm for MCS selection.
	\item DL/CA+I\textsc{wl}: Here, a deep-learning-based algorithm is applied for channel access, and the Intel I\textsc{wl} algorithm is used for MCS selection.
\end{itemize}

It should be noted that both algorithms, $''$CSMA/CA+DL/MCS$''$ and $''$DL/CA+I\textsc{wl}$''$, are subject to the half-duplex constraint of radio hardware, which prevents AI stations from sensing the channel while transmitting. Consequently, it is necessary to conduct RSSI compensation operations after transmission has finished. To ensure a fair and accurate comparison with our proposed DL-MAC algorithm, both strategies  implement RSSI compensation in accordance with the same principles outlined in Section \ref{subsec:imp_jcara}.

Additionally, the RTS/CTS control frames serve as an effective channel reservation mechanism that can significantly enhance network performance in Wi-Fi networks. Because RTS/CTS frames are short, reducing the likelihood of collisions, and even when collisions occur, the associated cost is minimal. However, it is important to clarify that our approach, which is based on RSSI data for channel access, differs in its objectives from the RTS/CTS mechanism. Our method aims to avoid potential conflicts with existing devices in the network by accessing idle channels, whereas RTS/CTS frames are primarily designed to address conflicts during transmission. In our wireless networks, due to the need for flexible utilization of unoccupied channels, the traditional RTS/CTS control frame mechanism is not entirely suitable, as it may interfere with existing network devices. Therefore, we do not adopt the RTS/CTS mechanism as a baseline for comparison.

\subsection{ Performance evaluation on single channel}
In this set of experiments, we designate channel 6 as the operational channel for DL-MAC. Initially, we employ offline training for the RNN, utilizing data collected from the first 60 seconds on this channel. The dataset is divided into two subsets: 80\% is allocated as the training set, and the remaining 20\% forms the validation set. To mitigate the risk of overfitting, we employ early stopping, a technique that halts training when there is no observed improvement in the model$'$s performance on the validation set over a specified number of epochs. Following this offline training phase, we conduct online testing to assess the performance of DL-MAC using fresh data from the subsequent 10 seconds. Each experiment is conducted 10 times, with the mean and standard deviation of outcomes reported. It should be noted that the training process is conducted in an offline manner, and the testing process is conducted online.

\subsubsection{Performance comparison of different algorithms}
We first compare the performance of DL-MAC against that of traditional algorithms in terms of throughput and mean delay, using OPT as the benchmark for the upper bound of MAC performance. The comparative results for throughput and mean delay are depicted in Fig. ~\ref{fig:ch6_performance-comparison_a} and Fig. ~\ref{fig:ch6_performance-comparison_b}, respectively. Among the protocols compared, OPT consistently outperforms others, achieving the highest throughput and the lowest delay across all time intervals. The traditional algorithms, specifically CSMA/CA+I\textsc{wl} and CSMA/CA+ARF, not only exhibit lower throughput and higher packet delays but also demonstrate considerable performance fluctuations over different time intervals.

In contrast, DL-MAC presents a significantly better throughput than traditional algorithms. Similarly, Fig. ~\ref{fig:ch6_performance-comparison_b} indicates that DL-MAC incurs a mean delay that, while higher than that of the OPT, is substantially lower than that of the traditional algorithms. The performance of DL-MAC, in terms of both throughput and delay, not only outperforms traditional algorithms but also demonstrates stability and minimal fluctuation, closely approaching the optimal performance benchmark.

 \subsubsection{Generalization performance}
 Fig. \ref{fig:generalization-comparison} displays the generalization performance of the DL-MAC model, which was trained on channel 6 and then tested on channels 1 and 11. The figure also presents two performance metrics-throughput and mean delay-compared against the OPT benchmarks and the CSMA/CA+I\textsc{wl} algorithm for each respective channel.
 
Across both tested channels, DL-MAC demonstrates better generalization performance than CSMA/CA+I\textsc{wl}, as indicated by higher throughput and lower mean delay. The performance of DL-MAC is closer to that of OPT, showcasing the effectiveness of DL-MAC in adapting to different channel conditions. Moreover, the traditional algorithm CSMA/CA+I\textsc{wl} displays performance fluctuations across various time intervals on both channels 1 and 11, echoing the variability previously noted on channel 6. In summary, Fig. ~\ref{fig:generalization-comparison} demonstrates that DL-MAC has strong generalization capabilities across channels 1 and channel 11, approaching the optimal and surpassing traditional algorithms. This consistent effectiveness, despite the dynamic nature of network conditions, underscores the model$'$s reliable performance and its versatility in handling variations in channel characteristics.

\begin{figure}[!t]
  \centering
  \begin{subfigure}{0.85\linewidth}
    \includegraphics[width=\linewidth]{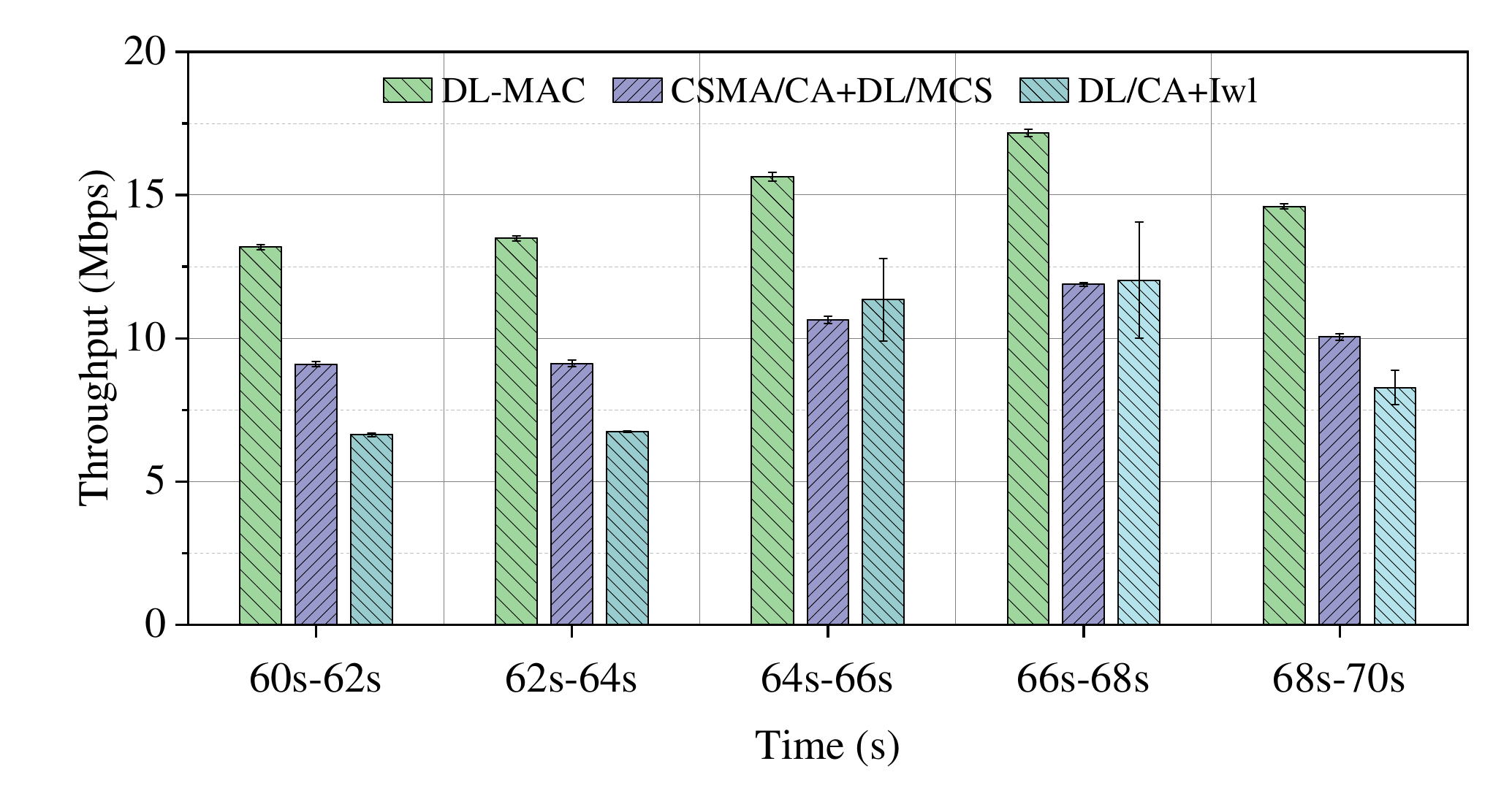}
    \subcaption{Throughput}
    \label{fig:jointVSindi_a}
  \end{subfigure}
  
  \begin{subfigure}{0.85\linewidth}
    \includegraphics[width=\linewidth]{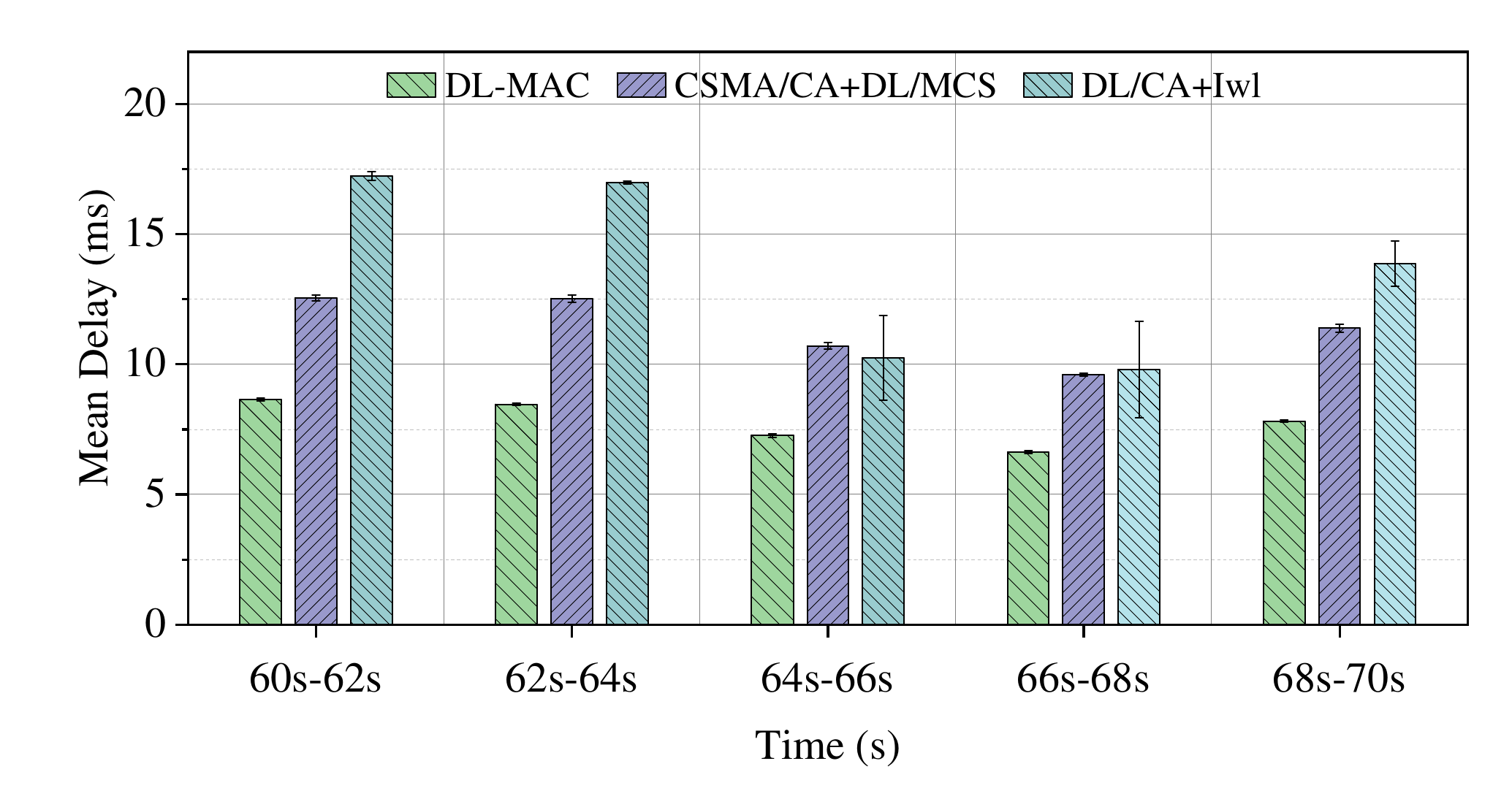}
    \subcaption{Mean delay}
    \label{fig:jointVSindi_b}
  \end{subfigure}
  
  \caption{Performance comparison of joint design vs. individual designs: (a) Throughput; (b) Mean delay.}
  \label{fig:jointVSindi}
\end{figure}

 \subsubsection{Joint design versus individual design}
 Now, we conduct a performance comparison, evaluating the joint design (i.e., the joint design of channel access and rate adaptation selection) against the deep-learning-based channel access (DL/CA+I\textsc{wl}) algorithm and the deep-learning-based MCS selection (CSMA/CA+DL/MCS) algorithm when designed separately. This is to verify whether our proposed joint design exhibits a performance advantage against individual designs.
 
As shown in Fig. \ref{fig:jointVSindi_a}, DL-MAC shows higher throughput compared to both CSMA/CA+DL/MCS and DL/CA+I\textsc{wl} across all time intervals, indicating that the joint design may be more effective in managing network resources for data transmission. The individual designs display lower throughput compared to the joint design, suggesting that the joint design of channel access and MCS selection may provide synergistic benefits that are not fully realized when these functions are separated. From Fig. \ref{fig:jointVSindi_b}, the mean delay is lowest for DL-MAC, which aligns with its higher throughput, suggesting efficient packet handling and reduced waiting times. Both individual design approaches, incur higher mean delays. This further supports the notion that the joint design approach can achieve faster packet transmission. This comparison indicates that the proposed joint design paradigm is effective, potentially due to better-coordinated decision-making processes that take into account the interdependencies between channel access and MCS selection.

\begin{figure}[!t]
  \centering
  \begin{subfigure}{0.85\linewidth}
    \includegraphics[width=\linewidth]{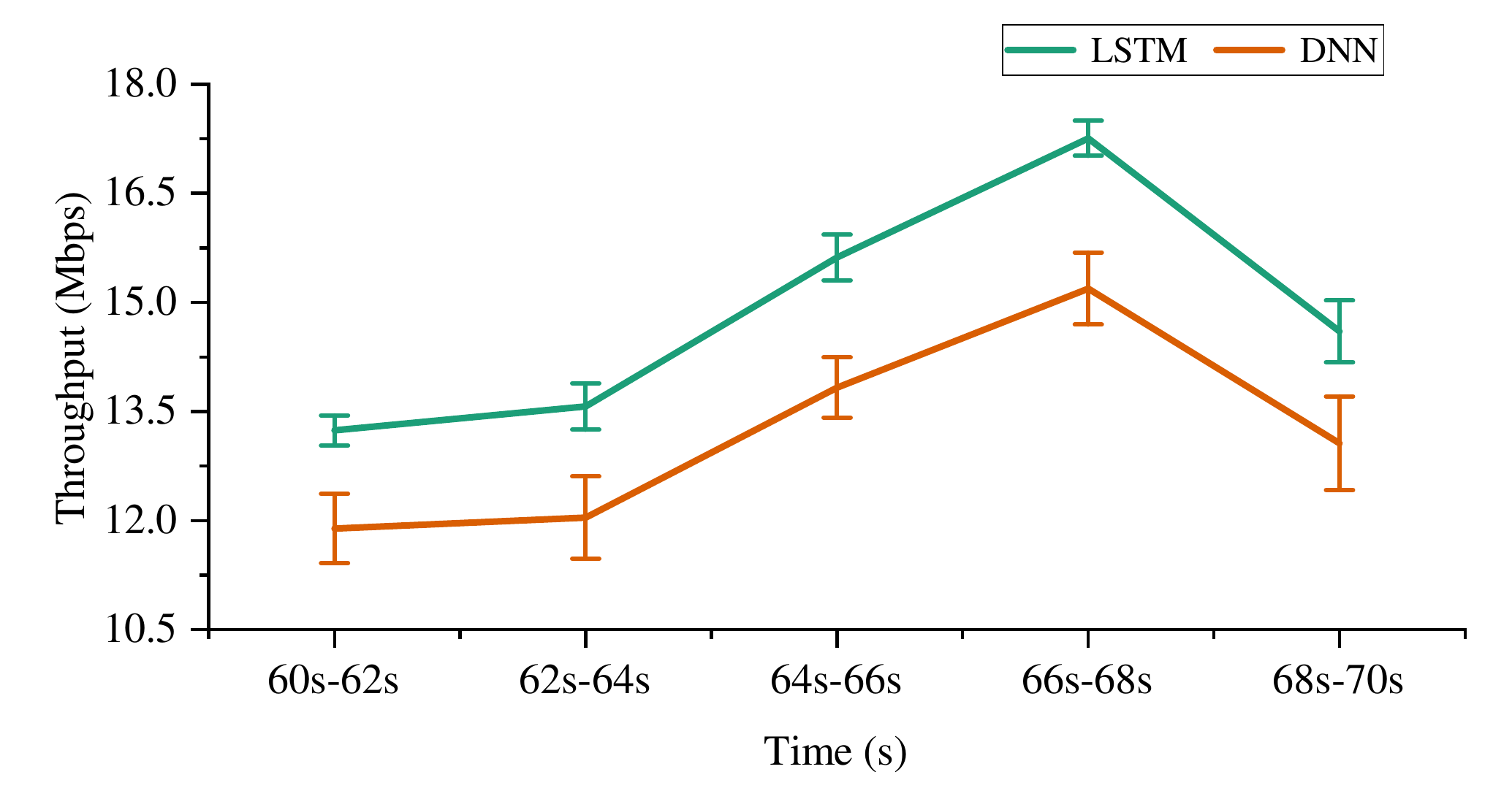}
    \subcaption{Throughput}
    \label{fig:lstmVSdnn_a}
  \end{subfigure}
  
  \begin{subfigure}{0.85\linewidth}
    \includegraphics[width=\linewidth]{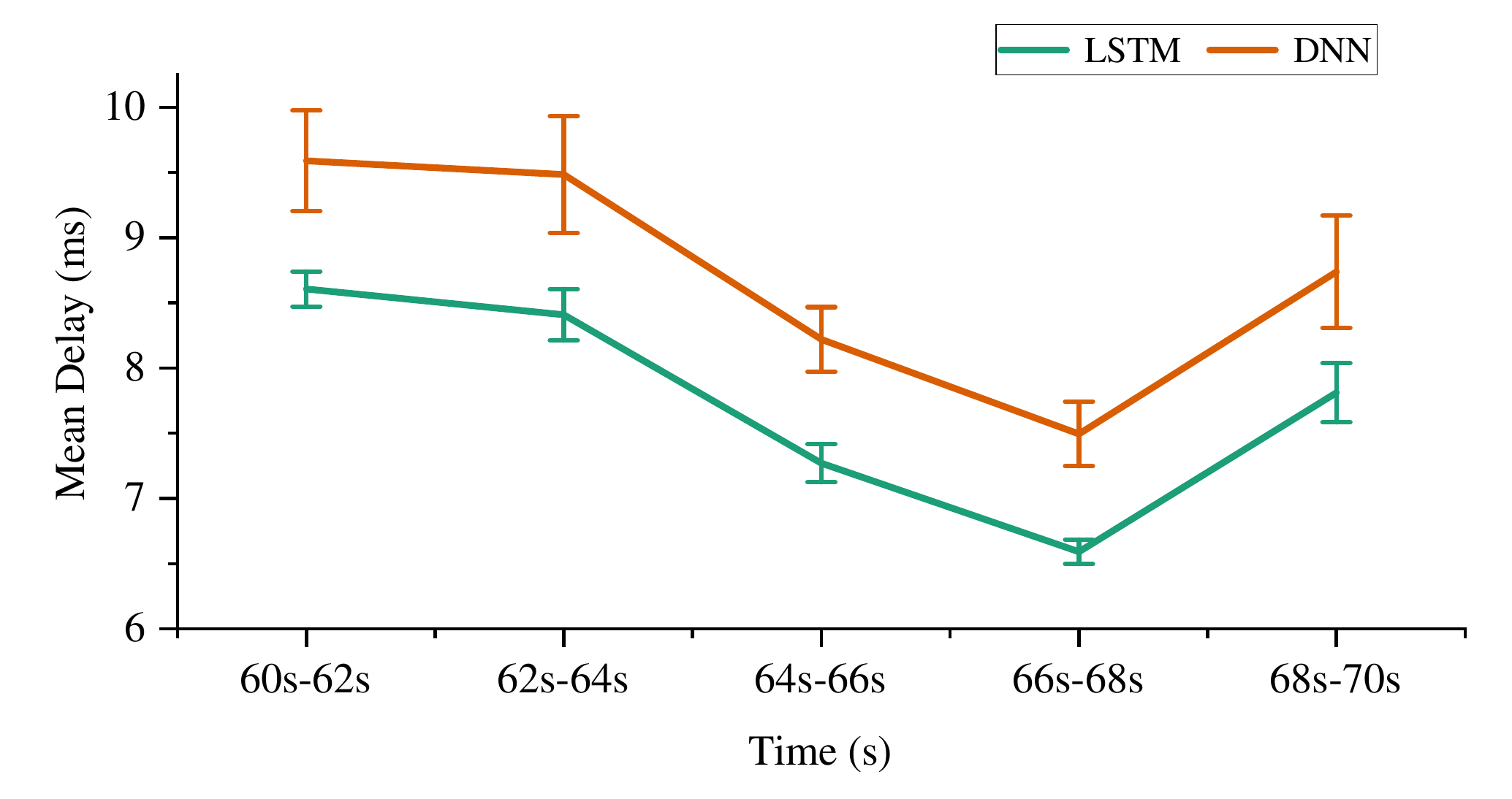}
    \subcaption{Mean delay}
    \label{fig:lstmVSdnn_b}
  \end{subfigure}
  
  \caption{Comparing the performance of schemes using LSTM and plain DNN: (a) Throughput; (b) Mean delay.}
  \label{fig:lstmVSdnn}
\end{figure}

\subsubsection{LSTM versus plain DNN}
We proceed to investigate the impact of different neural network architectures on the performance of the joint design of channel access and rate adaptation within DL-MAC. Specifically, we examine two architectures: LSTM, as previously proposed, and a plain DNN. For the plain DNN, we just employ a straightforward fully connected network with three hidden layers, which consist of 512, 128, and 64 neurons, respectively. The ReLU function is employed as the activation function for the hidden neurons. Fig. ~\ref{fig:lstmVSdnn} presents a comparison of these two schemes, focusing on their performance in terms of throughput and mean delay. The error bars represent three times the standard deviation, derived from the mean and variance obtained over ten experiments.

As we can see from Fig. ~\ref{fig:lstmVSdnn_a}, the LSTM model consistently achieves higher throughput across the measured time intervals than the plain DNN model. This is attributed to the LSTM$'$s inherent suitability for processing time series data, enabling it to handle RSSI sequences more effectively for DL-MAC applications. Moreover, observed in Fig. ~\ref{fig:lstmVSdnn_b}, the LSTM model displays a lower mean delay, indicating its potential for tasks that demand both high throughput and low latency within DL-MAC$'$s framework for joint channel access and rate adaptation. On average, the LSTM model achieves approximately 12.5\% higher throughput and an 11.2\% reduction in mean delay compared to the plain DNN model, highlighting its superior ability to capture temporal dependencies in RSSI data. This enables the LSTM to more effectively optimize the task of joint channel access and rate adaptation based on a series of RSSI data during online testing.

\begin{figure}[!t]
  \centering
  \begin{subfigure}{0.85\linewidth}
    \includegraphics[width=\linewidth]{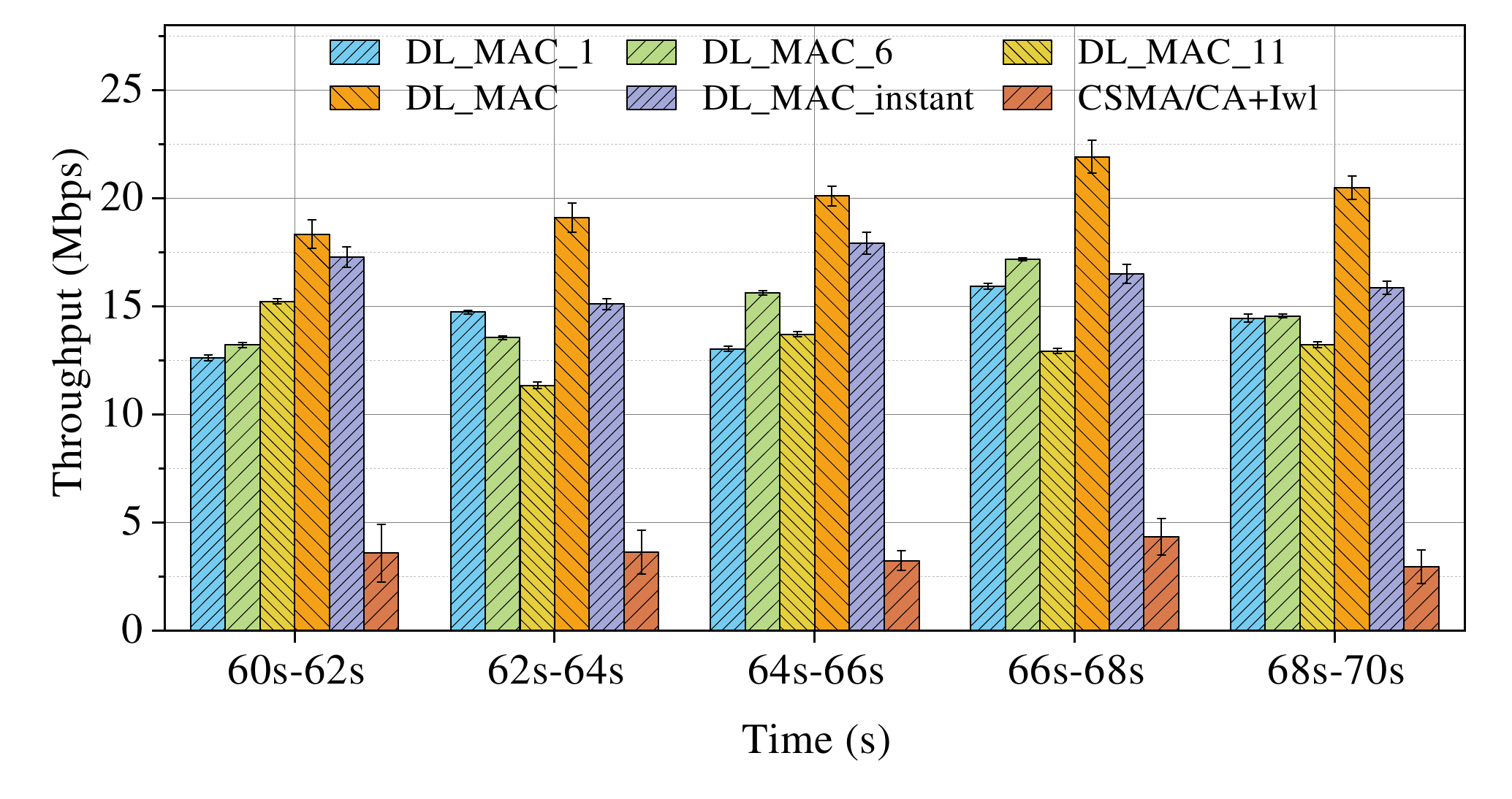}
    \subcaption{Throughput}
    \label{fig:channel_switch_a}
  \end{subfigure}
  
  \begin{subfigure}{0.85\linewidth}
    \includegraphics[width=\linewidth]{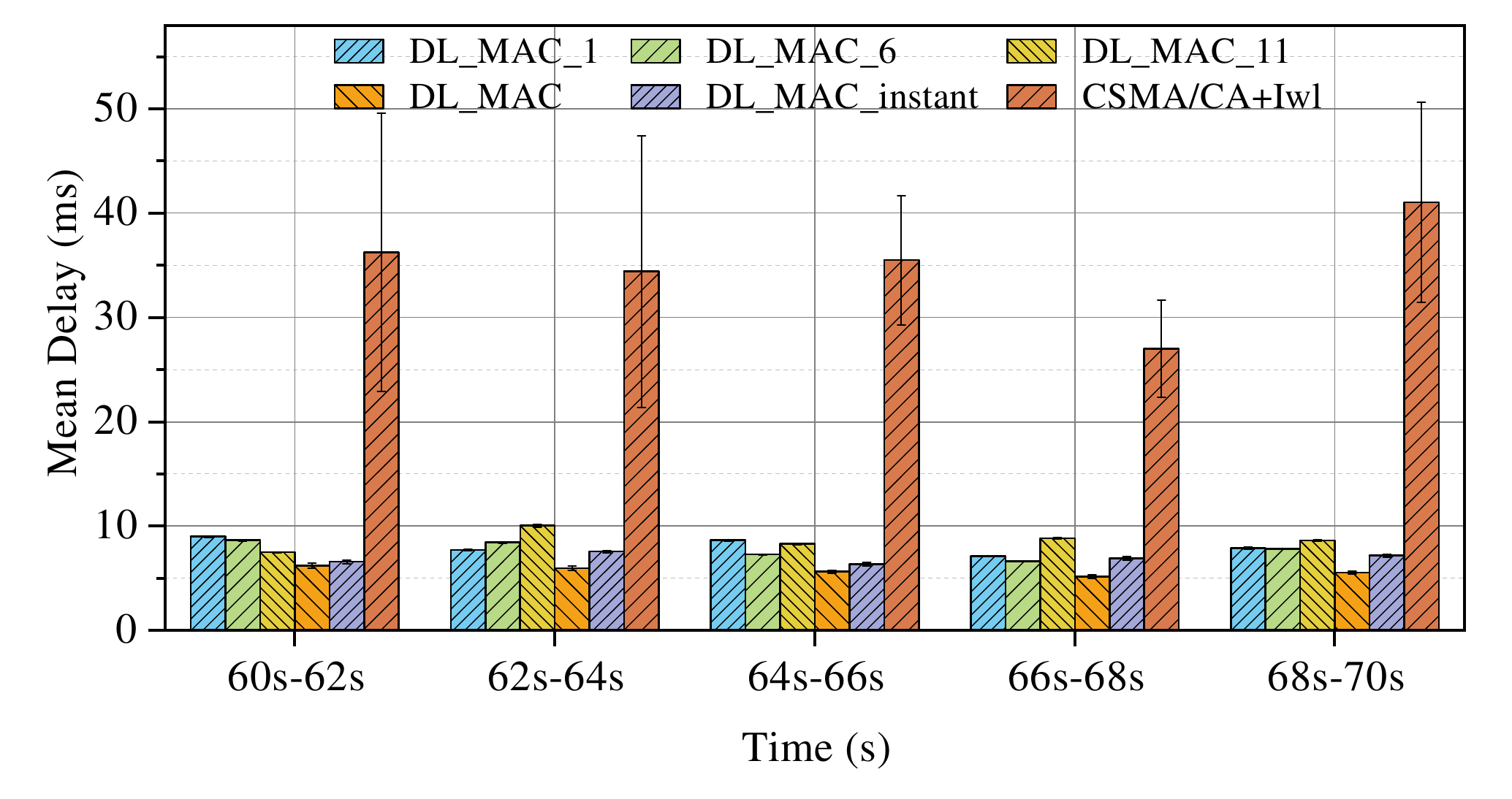}
    \subcaption{Mean delay}
    \label{fig:channel_switch_b}
  \end{subfigure}
  
  \caption{Evaluating channel switch performance in DL-MAC protocol: (a) Throughput; (b) Mean delay.}
  \label{fig:channel_switch}
\end{figure}

\subsection{Performance evaluation across multiple channels}
In this set of experiments, we designate channels 1, 6, and 11 as the available options for AI device operations. This configuration allows the DL-MAC protocol to dynamically select the optimal channel from among these channels for channel access and rate adaptation at varying time intervals. For the DNN training, we utilize data collected from these three channels during the first 60 seconds. We  then partition the dataset into training and validation subsets, allocating 80\% to training and the remaining 20\% to validation. An early stopping mechanism is implemented to prevent overfitting during the offline training process. Once  trained, the DNN is utilized online to perform the channel switch process. The performance of DL-MAC is then evaluated using a separate set of data from the following 10 seconds. Likewise, each experiment is replicated 10 times, and we report the mean and standard deviation for all assessed metrics. Likewise, the training process is conducted offline, and the testing process is conducted online.

\subsubsection{Performance comparison of channel switch}
Fig. \ref{fig:channel_switch} displays the performance of the DL-MAC protocol with channel switch capabilities across channels 1, 6, and 11. In the figure, the performance of the DL-MAC protocol operating on channels 1, 6, and 11 is denoted as DL-MAC\_1, DL-MAC\_6, and DL-MAC\_11, respectively. Additionally, we have introduced an instant channel switch algorithm named $''$DL-MAC\_instant$''$. This algorithm performs real-time predictions of all available channels immediately after each transmission,  selecting the optimal channel for the next channel access and transmission. This achieves real-time optimization of channel selection. Furthermore, we have compared this with the traditional algorithm using the CSMA/CA protocol for channel access combined with the Intel I\textsc{wl} algorithm for rate adaptation.

The results presented in Fig. \ref{fig:channel_switch} demonstrate the performance advantages of the DL-MAC protocol with channel switch capabilities. We have three main observations from Fig. \ref{fig:channel_switch}:

\begin{itemize}
	\item Our proposed DL-MAC protocol with channel switch capability demonstrates higher throughput and lower delay in multi-channel environments compared to when DL-MAC operates on a single channel. Despite the inherent overhead associated with the channel switch, the overall performance improvement validates the effectiveness of the channel switch capability. The impact of channel switch overhead has been experimentally verified in subsection \ref{subsec:impact_cs_cost}.
	\item The instant channel switch algorithm, although capable of rapidly adapting to changes in the network environment and achieving good throughput and delay performance, still falls short of the DL-MAC with integrated channel switch capability. The instant channel switch algorithm can predict and select the best channel immediately after each transmission. However, due to rapid changes in channel conditions and frequent switching, it ultimately leads to reduced throughput and increased latency in practical applications.
	\item For device employing the CSMA/CA protocol for channel access coupled with the Intel I\textsc{wl} algorithm for rate adaptation, there is a significant reduction in throughput as well as an increase in mean delay. The primary reason for this performance degradation is the reliance of the I\textsc{wl} algorithm on historical information from previous channel conditions for rate decision-making. Consequently, when device transition to a new channel, the lack of real-time detection and evaluation of the new channel conditions impacts the accuracy of rate adaptation. Often, the algorithm may select a rate that is below the optimal or feasible rate due to the lack of current channel data. This situation can severely degrade network performance, especially in environments with frequent channel switch.
\end{itemize}

\begin{figure}[!t]
  \centering
  \begin{subfigure}{0.85\linewidth}
    \includegraphics[width=\linewidth]{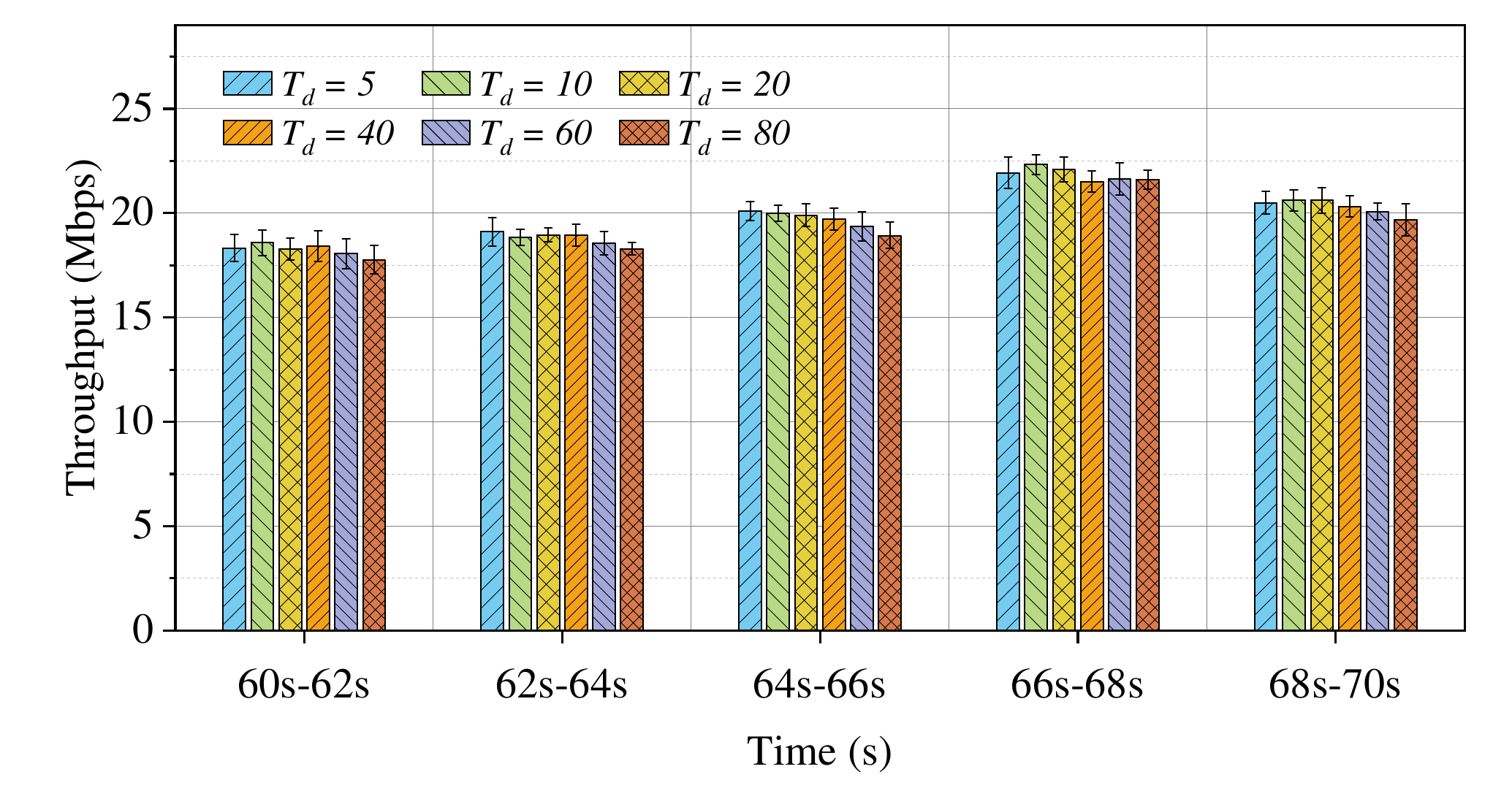}
    \subcaption{Throughput}
    \label{fig:switch_cost_a}
  \end{subfigure}
  
  \begin{subfigure}{0.85\linewidth}
    \includegraphics[width=\linewidth]{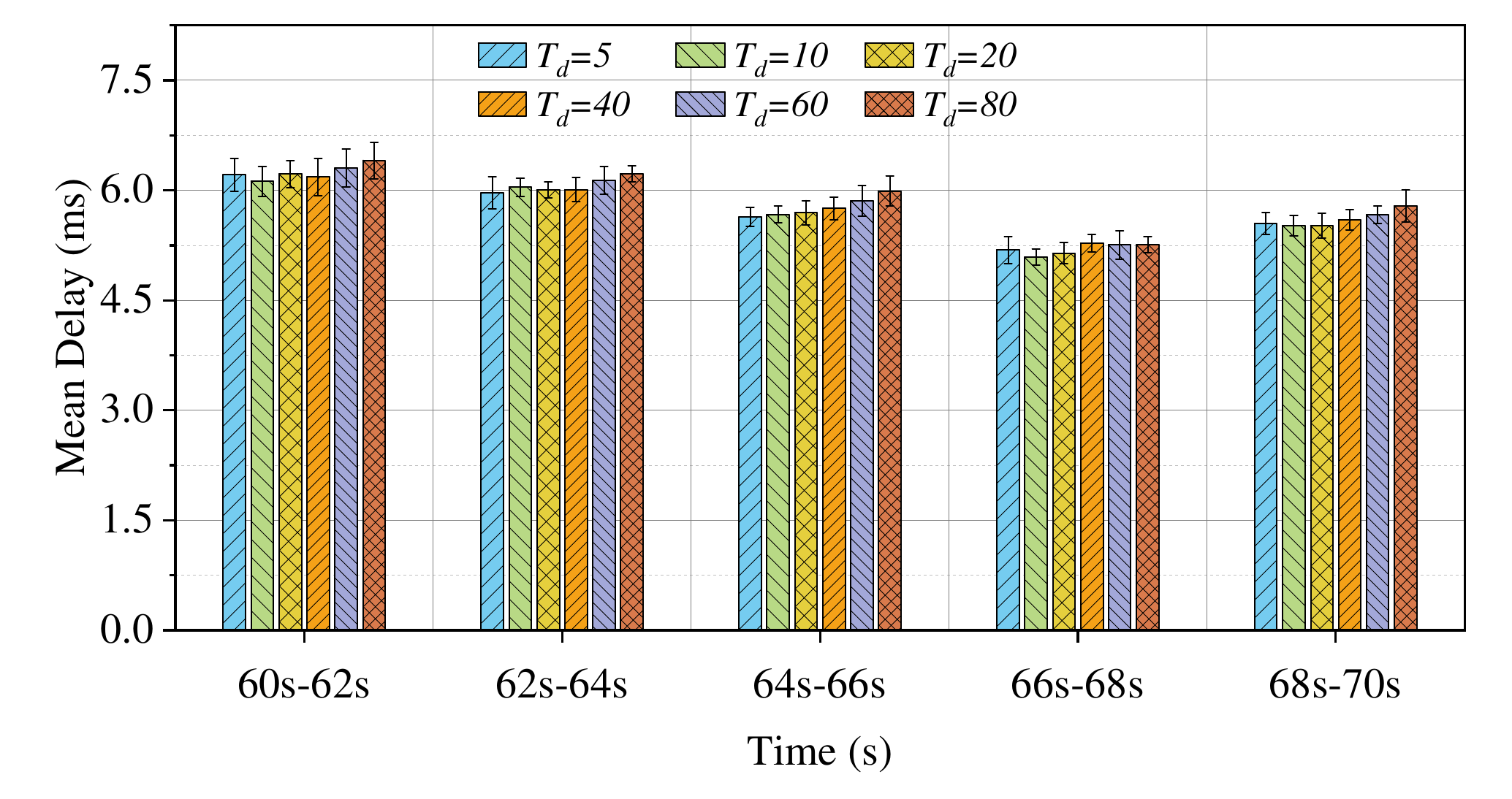}
    \subcaption{Mean delay}
    \label{fig:switch_cost_b}
  \end{subfigure}
  
  \caption{Performance comparison under different channel switch overhead ($T_d$), showing: (a) Throughput; (b) Mean delay.}
  \label{fig:switch_cost}
\end{figure}

\subsubsection{Impact of the channel switch overhead}\label{subsec:impact_cs_cost}
We next examine the impact of channel switch overhead, denoted by $T_d$, on the performance of DL-MAC. Fig. \ref{fig:switch_cost} illustrates the performance comparison under varying channel switch overheads. As observed from Fig. \ref{fig:switch_cost_a}, the throughput remains relatively consistent across a range of $T_d$ values, from $T_d=5$ to $T_d=80$. There is a negligible or no significant reduction in throughput as $T_d$ increases, suggesting that DL-MAC can effectively manage channel switch without a substantial impact on throughput within the tested range of $T_d$ values. Similar to throughput, the mean delay across different $T_d$ values remain relatively consistent, without noticeable increase as $T_d$ rises, as shown in Fig. \ref{fig:switch_cost_b}. This suggests that the protocol can maintain timely packet delivery even as the cost of channel switch varies.

In summary, Fig. \ref{fig:switch_cost} shows that the DL-MAC protocol$'$s performance, in terms of throughput and mean delay, is not significantly affected by the channel switch overhead within the tested range. This could indicate that DL-MAC is effectively managing the channel switch process, maintaining performance despite the incurred delays from switching. The experimental results suggest that the DL-MAC protocol exhibits resilience to the overheads introduced by channel switch, which is a desirable feature in a dynamic multi-channel environment.

\begin{figure}[!t]
  \centering
  \begin{subfigure}{0.85\linewidth}
    \includegraphics[width=\linewidth]{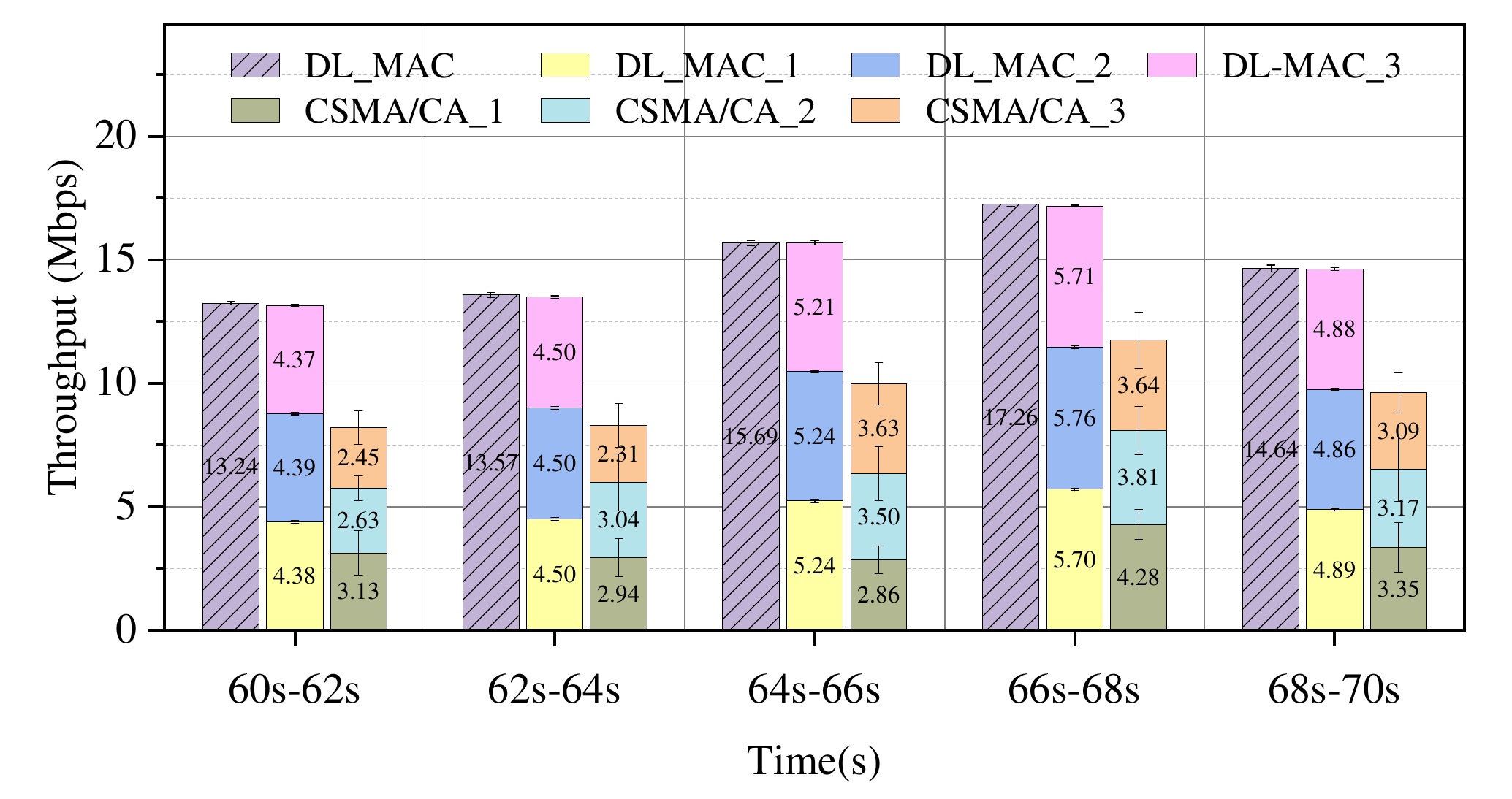}
    \subcaption{Throughput}
    \label{fig:threeSTAs_a}
  \end{subfigure}
  
  \begin{subfigure}{0.85\linewidth}
    \includegraphics[width=\linewidth]{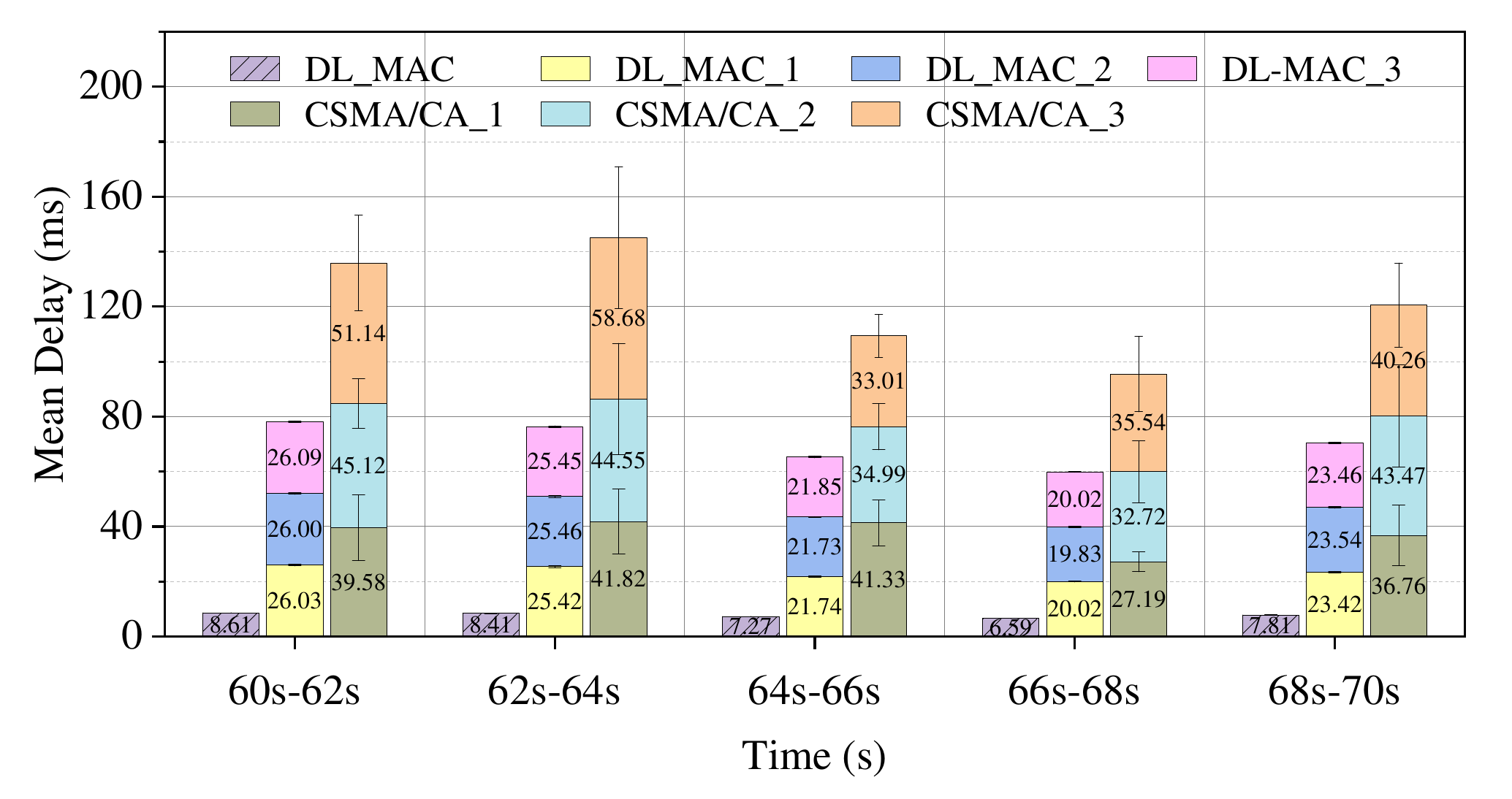}
    \subcaption{Mean delay}
    \label{fig:threeSTAs_b}
  \end{subfigure}
  
  \caption{Performance comparison of three DL-MAC devices in a unified protocol setting: (a) Throughput; (b) Mean delay.}
  \label{fig:threeSTAs}
\end{figure}

\subsection{DL-MAC reformulation}
Note that all of our previous experiments were conducted in an environment where only one intelligent device coexists with multiple legacy devices. However, when there are multiple intelligent devices in the DL-MAC network, the complexity of the system significantly increases due to all intelligent devices making independent decisions within the same environment. This is because each intelligent device can be considered an entity whose transmission behavior needs to be learned by the other intelligent devices.

To address this issue, we have reformulated our system model as a semi-distributed system. Within this system, all intelligent devices in the DL-MAC network are coordinated by a central DL-MAC gateway, akin to the point coordination function (PCF) protocol within the IEEE 802.11 MAC architecture. Under this model, the DL-MAC gateway is first subjected to offline training, a process consistent with that of a single intelligent DL-MAC device. Following the completion of the DL-MAC gateway$'$s training, it can then be employed for online testing. In each time slot, the DL-MAC gateway decides whether intelligent devices within the DL-MAC network should access the channel. If the gateway decides to allow channel access and transmission, it selects one intelligent device within the DL-MAC network to carry out the transmission via a polling mechanism. After transmission, the chosen DL-MAC intelligent device receives feedback from the system and communicates with the gateway, which then performs the corresponding RSSI compensation. If the gateway decides not to allow channel access, then all intelligent devices remain silent. In this manner, the DL-MAC gateway can essentially be regarded as a large virtual proxy composed of multiple DL-MAC intelligent devices combined. As for the coordination information between the gateway and the intelligent devices it manages, this can be achieved through signal transmission over the control channel, which is typically completed within a shorter time slot. In this paper, we do not delve into this part in detail because our focus is primarily on the model architecture rather than the implementation specifics.

In our experimental setup, a DL-MAC gateway includes three DL-MAC intelligent devices, all of which operate in a saturated traffic mode, meaning each device always has packets to send. Each device is equipped with a buffer to store the packets awaiting transmission, and the arrival process of these packets follows a Poisson distribution, as described in Section \ref{sect:system parameters}. Unless specifically stated otherwise, all experimental settings remain consistent with our previous experiments.

Fig. \ref{fig:threeSTAs} displays the performance of throughput and mean delay of the DL-MAC gateway and that of the three DL-MAC intelligent devices it manages. We use the performance of the DL-MAC gateway as the benchmark, considering that in heterogeneous network scenarios, the DL-MAC gateway can be viewed as an independent DL-MAC device.  For comparison to our proposed DL-MAC, we also include three devices that employ the traditional CSMA/CA protocol for channel access and the Intel I\textsc{wl} algorithm for rate selection. In Fig. \ref{fig:threeSTAs}, DL-MAC is represented as the gateway proxy, while DL-MAC\_1, DL-MAC\_2, and DL-MAC\_3 are the three intelligent devices within the DL-MAC gateway, respectively; CSMA/CA\_1, CSMA/CA\_2, and CSMA/CA\_3 represent the devices that adopt the CSMA/CA protocol for channel access and the Intel I\textsc{wl} algorithm for rate selection, respectively. From the results displayed in Fig. \ref{fig:threeSTAs_a}, it is evident that the combined throughput of the intelligent devices within the DL-MAC gateway, which employs a polling mechanism, can achieve performance levels comparable to that of a single DL-MAC gateway. This not only reflects the system$'$s ability to integrate performance but also ensures fairness among the intelligent devices inside the gateway, with each device having equal opportunities for transmission. In contrast, devices utilizing the CSMA/CA protocol exhibit lower total throughput, and there are significant differences in fairness between devices, failing to guarantee equal channel access opportunities for each device. As we can see in Fig. \ref{fig:threeSTAs_b}, it is clear that the DL-MAC gateway exhibits the best performance in terms of mean delay. It can be inferred that having only a single DL-MAC device configured within the DL-MAC gateway allows this device to have the opportunity for channel access and transmission in every communication cycle, thereby achieving the lowest delay. The men delay of the three DL-MAC devices within the gateway is approximately three times that of the single device within the DL-MAC gateway, a result directly attributed to the operation of the polling mechanism, where each device takes turns accessing the channel and transmitting. Even so, compared to devices utilizing the CSMA/CA protocol, DL-MAC devices still demonstrate a significant advantage in mean delay under the polling mechanism. Moreover, devices within the DL-MAC gateway maintain uniform delay times, with no device experiencing unreasonable delays, showcasing the system$'$s fairness in delay management. For CSMA/CA devices, there is a notable variation in delay among devices. Therefore, the DL-MAC gateway and its management devices exhibit significant advantages in ensuring high throughput, low mean delay and maintaining fairness, further proving the effectiveness and reliability of the DL-MAC protocol in handling high-density network communications.

\section{Conclusion}\label{sect:conclusion}
In this paper, we have proposed a DL-MAC protocol, an innovative cross-layer decision-making strategy that incorporates deep learning techniques. This protocol integrates functionalities for channel switch and joint channel access and rate adaptation, aiming to optimize throughput and minimize packet delay in complex and heterogeneous networks. The DL-MAC framework utilizes a deep neural network (DNN) for precise channel selection and employs a recurrent neural network (RNN) to efficiently manage joint channel access and rate adaptation on the selected channel. By integrating these functions into a unified system, the DL-MAC achieves a multifunctional MAC protocol with a cross-layer design, significantly improving spectrum utilization and network performance. Experimental results from datasets collected in our laboratory demonstrate that the DL-MAC exhibits significant performance improvements in terms of throughput and latency compared to traditional MAC protocols, clearly proving the effectiveness of our proposed strategy in enhancing wireless network performance.

In future work, we aim to investigate effective methodologies for characterizing spectrum data sampled from the real world to  facilitate deeper data analysis. This research will focus on developing new data processing and analysis techniques to handle and utilize a large amount of real-world spectrum data, providing guidance for optimizing and improving the performance of wireless communication systems.

Another possible research direction involves designing more efficient algorithms to further optimize network throughput and minimize latency while considering fairness across the network. We will also investigate the joint design of channel access and transmission power control strategies. Currently, the DL-MAC protocol primarily focuses on channel access and rate adaptation. Incorporating transmission power control into the considerations can effectively manage the transmission strength of wireless signals, reducing interference, improving communication quality, and enhancing system energy efficiency in the future.

\section*{Declaration of competing interest}
The authors declare no conflict of interest.

\section*{Acknowledgments}
This work was supported by the National Key R\&D Program of China under Grant 2021YFB1714100, and by the Shenzhen Science and Technology Program, China under Grant JCYJ20220531101015033.

\bibliographystyle{elsarticle-num}
\balance
\bibliography{egbib}







\end{document}